Accepted in *The Astrophysical Journal*, July 6, 2009

# THE 2008 AUGUST 1 ECLIPSE SOLAR-MINIMUM CORONA UNRAVELED

J. M. PASACHOFF[1,2], V. RUŠIN[3], M. DRUCKMÜLLER[4], P. ANIOL[5],
M. SANIGA[3] AND M. MINAROVJECH[3]

[1] Department of Planetary Sciences, California Institute of Technology 150-21, Pasadena, CA 91126
[2] Williams College—Hopkins Observatory, Williamstown, MA 01267, USA
[3] Astronomical Institute, Slovak Academy of Sciences, 059 60 Tatranská Lomnica, Slovak Republic
[4] Faculty of Mechanical Engineering, Brno University of Technology, 616 69 Brno, Czech Republic
[5] ASTELCO Systems GmbH, Fraunhoferstr. 14, D-82152 Martinsried, Germany

**Abstract**. We discuss results stemming from observations of the white-light and [Fe XIV] emission corona during the total eclipse of the Sun of 2008 August 1, in Mongolia (Altaj region) and in Russia (Akademgorodok, Novosibirsk, Siberia). Corresponding to the current extreme solar minimum, the white-light corona, visible up to 20 solar radii, was of a transient type with well-pronounced helmet streamers situated above a chain of prominences at position angles 48°, 130°, 241° and 322°. A variety of coronal holes, filled with a number of thin polar plumes, were seen around the poles. Furthering an original method of image processing, stars up to 12 magnitude, a Kreutz-group comet (C/2008 O1), and a coronal mass ejection (CME) were also detected, with the smallest resolvable structures being of, and at some places even less than, 1 arcsec. Differences, presumably motions, in the corona and prominences are seen even with the 19-min time difference between our sites. In addition to the high-resolution coronal images, which show the continuum corona (K-corona) that results from electron scattering of photospheric light, images of the overlapping green-emission-line (530.3 nm, [Fe XIV]) corona were obtained with the help of two narrow-passband filters (centered on the line itself and for the continuum in the vicinity of 529.1 nm, respectively), each with FWHM of 0.15 nm. Through solar observations, on whose scheduling and details we consulted, with the Solar and Heliospheric Observatory, Hinode's XRT and SOT, TRACE, and STEREO, as well as Wilcox Solar Observatory and SOHO/MDI magnetograms, we set our eclipse observations in the context of the current unusually low and prolonged solar minimum.

## 1. INTRODUCTION

Although space-borne observations of the Sun are, in general, commonly thought to be superior in quality compared to ground-based data, this is not fully the case with the solar corona. Not only can observations of the solar corona during total eclipses of the Sun provide data unavailable from any current satellites (SOHO, STEREO, TRACE and/or Hinode, as well as the pending Solar Dynamics Observatory, SDO), but some of them are unique for their kind (e.g., observing, with high resolution, the white-light corona within one solar radius above the solar limb, a region mostly inaccessible from SOHO or STEREO). This claim has been substantiated by our recent observations of the corona during the 2005 and 2006 total eclipses (Pasachoff et al. 2006; Pasachoff et al. 2007) and Wang et al. (2007) during the 2006 eclipse, which – thanks to a new method of data processing and analysis (Druckmüller et al. 2006) – revealed the complicated fine structure of the corona with unprecedented resolution. The 2008 August 1 total eclipse provided a unique opportunity to test this novel approach and to confirm the existence of new features in the corona, and confirm that these fine features are actually corona rather than artifacts of the image processing. In this article, we summarize the results obtained by analyzing observations carried out during this eclipse by the Czech-German-Slovak expedition based in Mongolia with some comparisons to eclipse observations from Siberian Russia, all compared with near-simultaneous satellite observations. Eclipse science has been reviewed by Pasachoff (2009a, 2009b).

## 2. OBSERVATIONS AND DATA ANALYSIS

Observations of the white-light and green-line emission corona during the 2008 August 1 total eclipse were carried out in Mongolia by equipment comprising a number of telephoto lenses*, ranging from 1250 mm to 200 mm in focal length, two of them with 0.3 nm narrow-passband filters at working temperature 45°C for the green-line corona observations. One of the filters was centered at the [Fe XIV] emission line at 530.3 nm and the other in the close neighborhood of 529.1 nm, both placed in front of the focal planes. This positioning enabled us to subtract from the light in 530.3 nm the contribution coming from the continuum corona to obtain the image of the "pure" green-line emission corona. One lens was equipped with an Hα filter with FWHM of 6 nm. Exposure times ranged from 8 s to 1/4000 s. 255 images were taken with digital cameras during the total eclipse and about 2000 calibration images were taken immediately after the end of totality. Cameras were controlled with special Linux (Fedora) software Multican written by Jindřich Nový, which was installed in notebook computers. All the objectives were placed on the same heavy high-precision parallactic mounting.

### 2.1 *Observing Site and Parameters*

Our main observing site, Bor Udzuur, was located at an altitude of 1223 meters above sea level about 23 km south-west of the small village of Altaj; its coordinates were 45° 23.251′ north latitude and 92° 06.837′ east longitude, almost at the central line of the eclipse (umbral depth 97.6%) (Espenak & Anderson 2007). The period of totality was 18:03:35 – 18:05:39 local time (i.e., 11:03:35 – 11:05:39 UT), its duration of 2:04 s being almost identical with its expected duration (Espenak & Anderson 2007). During the eclipse, the Sun was 22° above the horizon and observing conditions were excellent**.** Figure 1 depicts the resulting image in a logarithmic brightness scale composed with 25 images of the white-light corona with three overlays – two overlays pertain to different coronal regions and the third one is an overlay inset for the Moon – after the data were processed and enhanced in order to visualize low-contrast structures invisible to the human eye.

---

* In particular: The TMB Optical Co.'s apochromat (TMB APO) 105 mm f=620mm with Baader Flat-field corrector, effective focal length=1250 mm, Rubinar 10/1000 mm, Maksutov-Cassegrain 6.3/500 mm, 5.6/400 mm lens and 2.8/200 mm lens for the white-light corona. Two 500 mm, f/8 lenses for green-line emission and 65 mm ED APO f=400 mm for Hα. Cameras used were Canon EOS 1D Mark III, 1Ds Mark III, 5D, 20D, and 350D.

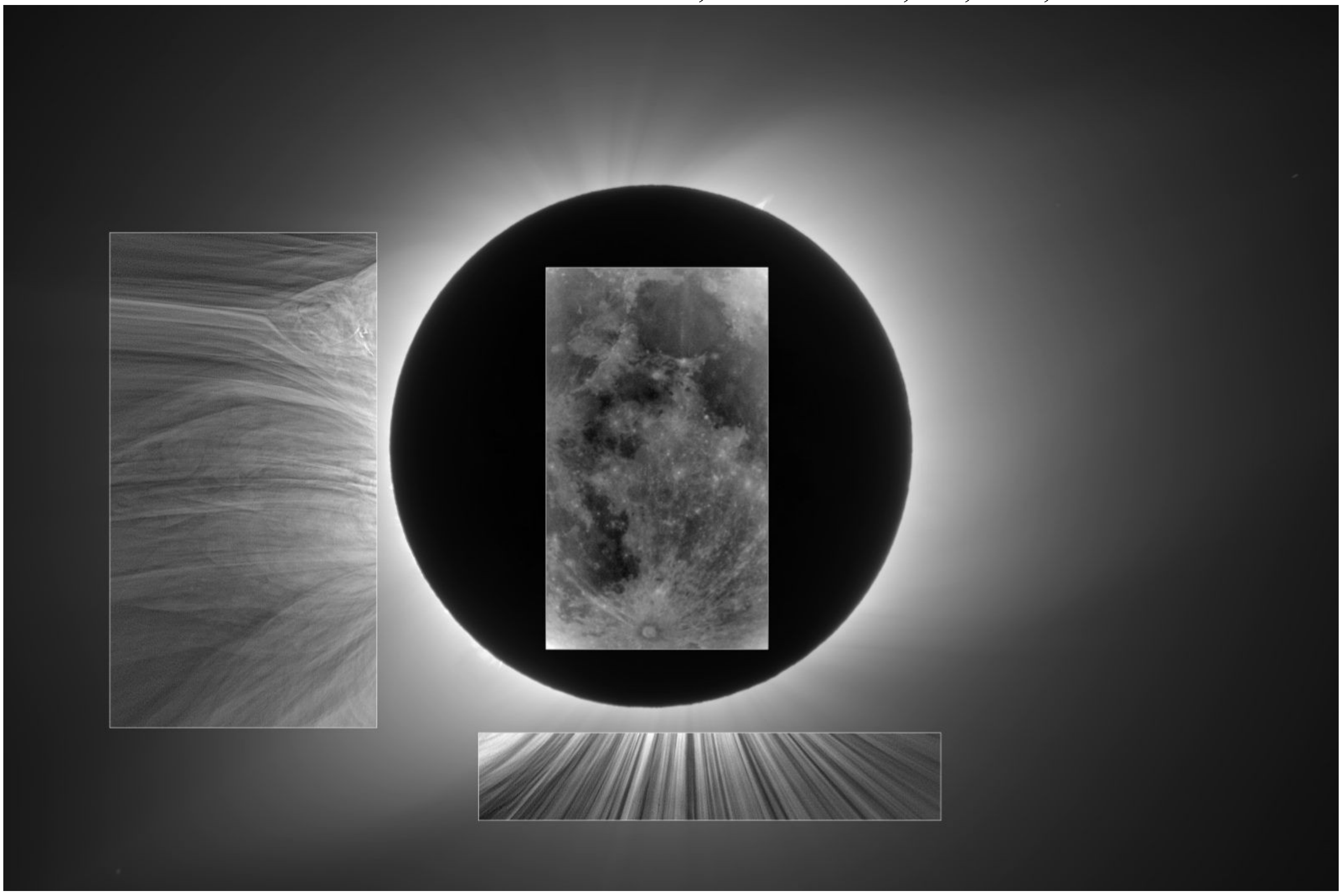

Fig. 1.— A logarithmic brightness scale image of the 2008 August 1 eclipse's white-light corona with insertions showing the corresponding parts after image processing. The pixel values of linear brightness scale image ranges by a factor of $10^7$. Enhanced images of the east limb's structure appear at left, of southern polar plumes appear at bottom, and of the lunar disk that is normally underexposured appear at center.

Processing the data obtained from such coordinated observations by Druckmüller’s method (Druckmüller et al. 2006) enabled us to get the images of the white-light corona with unprecedented ground-based resolution (see also Pasachoff et al. 2007 for the 2006 eclipse, including a basic discussion of the image-processing method, and Pasachoff et al. 2006 for the 2005 eclipse), of around or even slightly below 1 arcsec for observations carried out by the TMB Optical Company's apochromatic 105 mm lens and Baader Flat-field corrector with effective focal length 1250 mm. Over 100 plumes are visible in the enhanced image of the southern polar region, in an image covering about 1500 arcsec, making a spacing of about 15 arcsec. Individual features are finer, with some only 1 arcsec in width.

Note the details in the east-limb region, including overlapping structures that are not optically thick. Notice the turbulence above the prominence near the inset's top, and how the helmet, and thus the underlying magnetic field, converges to a straight spike. The extent of the non-radial structure is unprecedented.

2.2. *Details of Image Processing*

The raw images extracted from Canon CR2 files were at first calibrated by means of dark frames and flat-field images taken during several minutes after the totality. Four different dark frames and four different flat-field images were used for every eclipse image in order to minimize the additive noise. The calibrated images were aligned with sub-pixel precision. The alignment is based on coronal structures only in order to get maximum accuracy. The phase-correlation technique based on Fourier transform was used for image alignment. An unsharp mask was used to remove the effect of the lunar edge. Finally all individual images were composed in one single image in floating-point 3 x 64 bit pixel representation. The floating-point representation was used in order to minimize numerical errors. The method used for this eclipse was improved from that described by Druckmüller et al. (2006) and summarized by Pasachoff et al. (2007). Morgan, Habbal, & Woo (2006) and Habbal et al. (2008) have independently worked on high-resolution image processing, as has the group of Koutchmy (1997).

## 3. THE WHITE-LIGHT CORONA

We processed and analysed the white-light data separately for three distinct domains, differing from each other by increasing distance from the center of the Sun, in order to reveal extremely thin, small-scale structures of the inner corona as well as pronounced, large-scale objects at the most remote, still detectable outer parts of the corona. This strategy was dictated by a sharp drop in both the temperature and density of the corona with increasing distance from the Sun (e.g., Golub & Pasachoff 2009) and the technical capabilities/constraints of the observing equipment employed. Figure 2 shows the generic shape and fine structure of part of the corona above the eastern limb of the Sun. A careful inspection of the original images reveals a large number of super-fine objects/elements of diameter about or slightly less than 1 arcsec, i.e., about 700 km; such resolution has so far been reported only by Nesmyanovich et al. (1974) from observations of the eclipse in 1968 by a telescope whose focal length was 10 m, by Vsekhsvjatskyj et al. (1970), and by November & Koutchmy (1996), who went even below 1 arcsec thanks to excellent observing conditions and use of the 3.6 m Canada-France-Hawaii telescope at Hawai, the largest-diameter telescope ever to observe the corona. Alongside well-known loops at the feet/bases of helmet streamers and radial or slightly bent rays/plumes, our images also feature a number of bright and dark structures/elements of a helical shape which, to the best of our knowledge, are reported for the first time. The southwest limb of the Sun was covered by a number of small-scale, thin prominences.

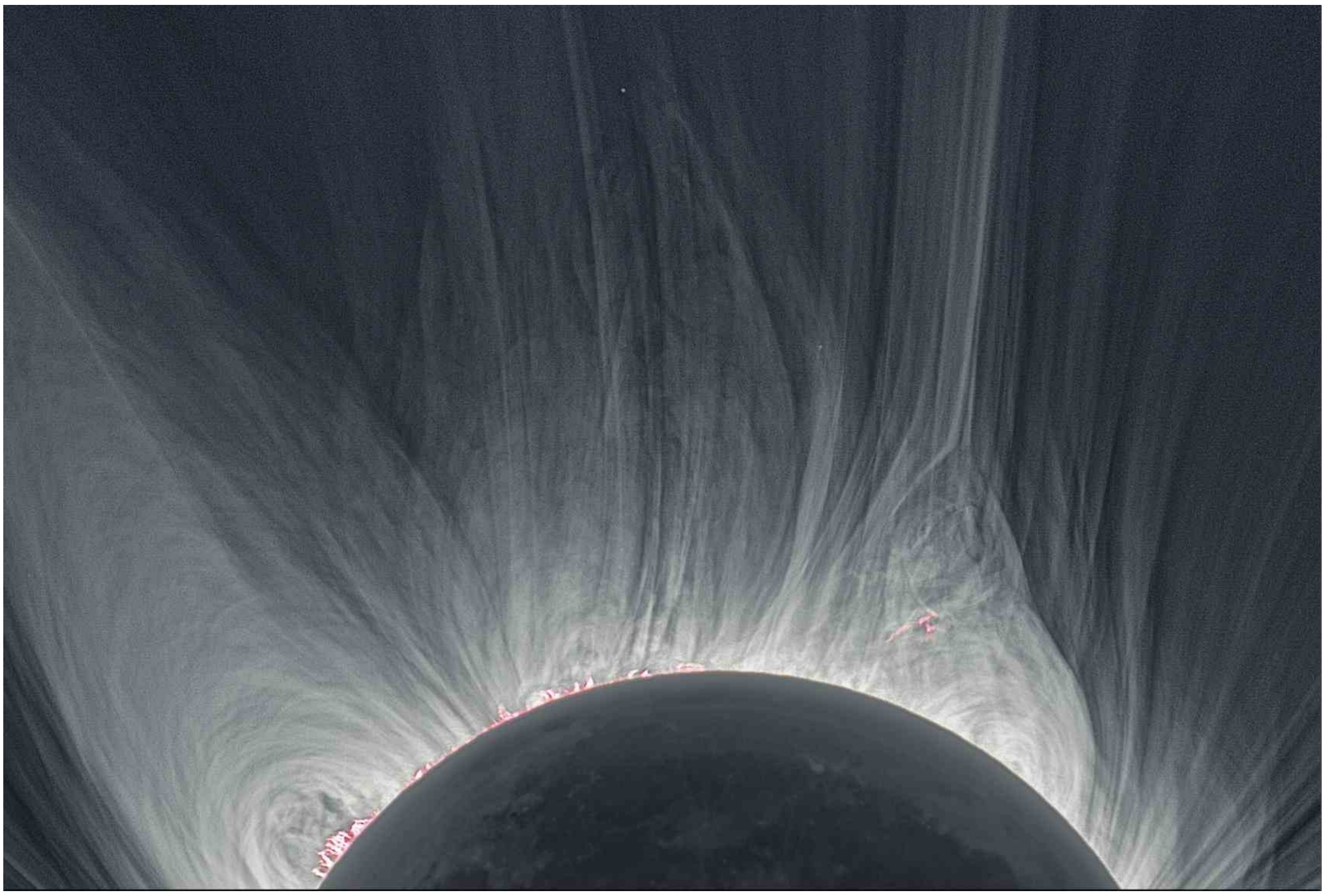

Fig. 2.— A processed composite image of the white-light corona above the eastern solar limb. Note the extent of the cavity associated with the prominences at left. (Lens: 105 mm TMB APO f=620mm + Baader Flat-field FFC EFL=1250 mm; 25 frames with exposure times of 1/4000 s to 8 s.)

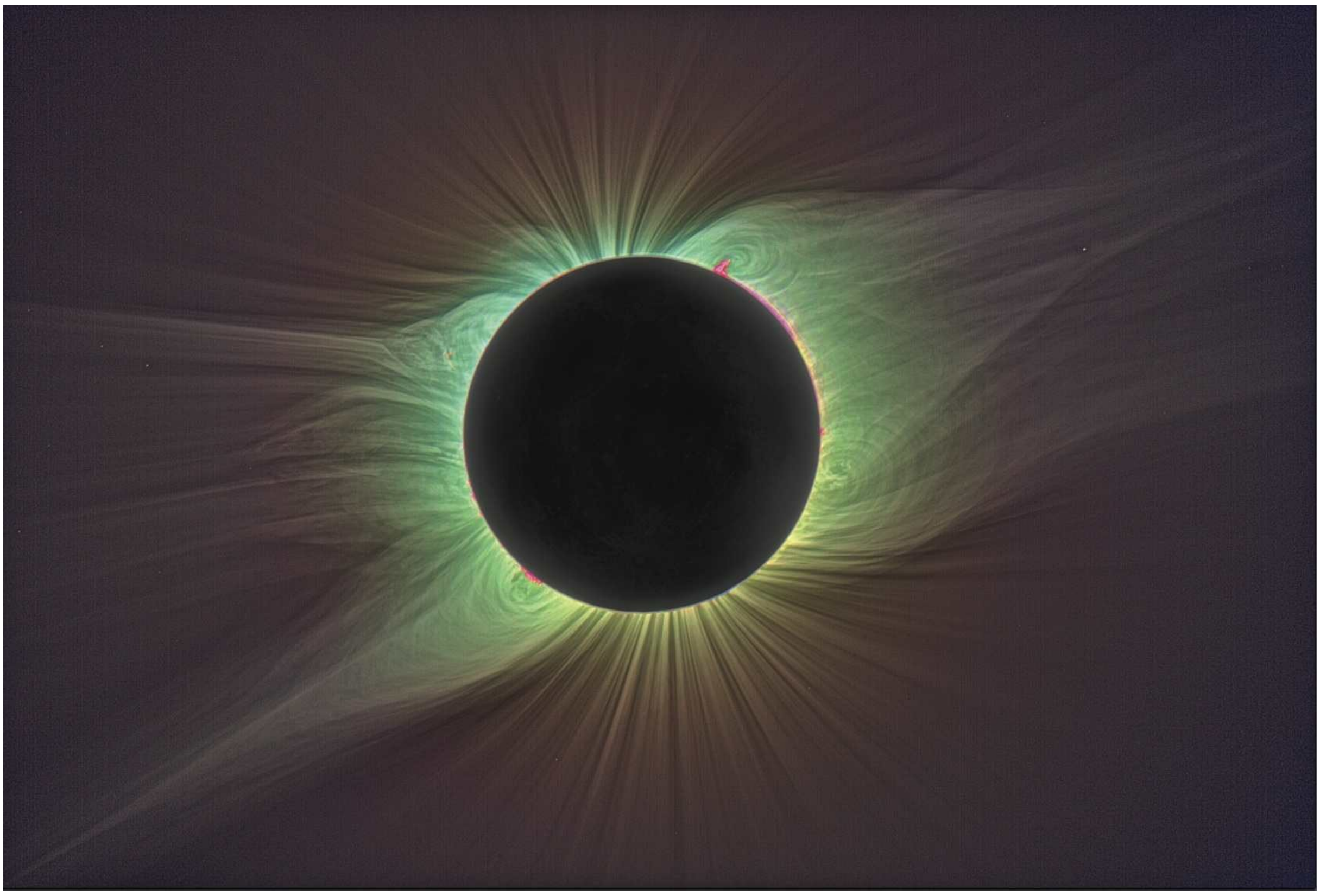

(a)

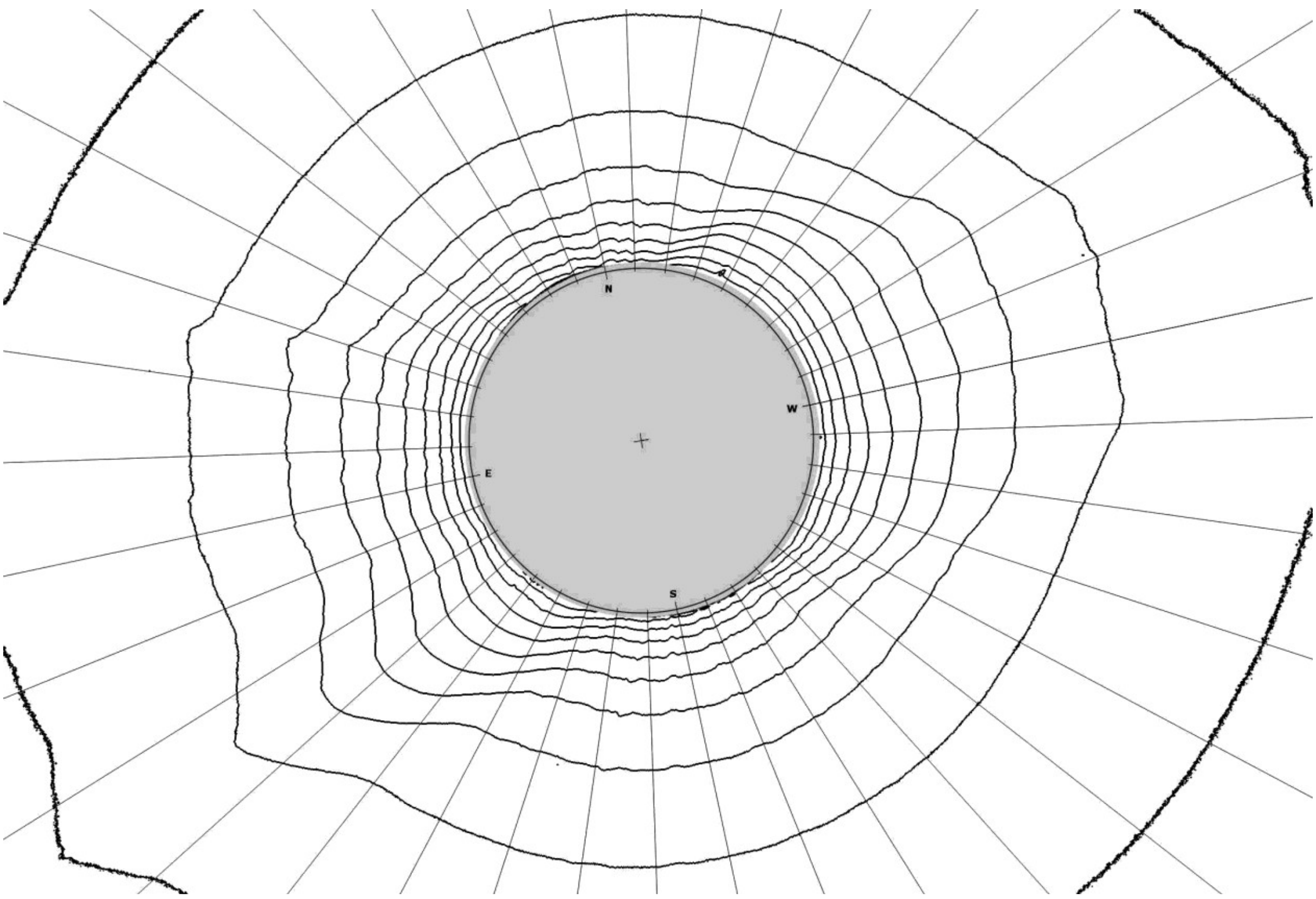

(b)

Fig. 3.— (a) A color-enhanced image (the saturation was increased by 50×) of the white-light corona inferred. (b) Isophotes on a logarithmic base 2; i.e., compared with the first isophote of arbitrary brightness 1, subsequent isophotes are for brightnesses reduced by 2, 4, 8, etc. (Data acquired with the Rubinar 10/1000 lens; 23 frames with exposures ranging from 1/60s to 8s.)

A color-enhanced composite of the white-light corona is given in Figure 3. To create the image, 23 exposures taken during the eclipse, and over two hundred calibration images – dark frames, flat-fields – were used to obtain precise color definition**.** Moreover, the blue color of the sky caused by Rayleigh scattering was removed. Finally the color saturation was increased 50×.

The image shows that this eclipse corona is of a solar-minimum type with well-developed polar plumes filling coronal holes around both the poles. Özkan et al. (2007) discuss the structure and flattening of the white-light corona at the 2006 eclipse; Golub and Pasachoff (2010) show an updated graph of the Ludendorff flattening coefficient through 2008. Ohgaito et al. (2002) show the falloff in intensity of the K- and F-coronas for the 1998 eclipse.

The feet of helmet streamers are localized on the channels of prominences mostly located at the positional angles of 48°, 130°, 241° and 322°, which is typical for the onset of a new solar cycle (see, e.g., Minarovjech 2007); large-scale structures pertinent to helmet streamers can be traced up to 20 solar radii (an estimate based on 55 exposures taken by 200 mm and 500 mm lenses with exposure times from 1/125 s to 8 s). Associated with coronal holes is a strange, low-contrast "curtain-like" structure, which we already reported in the 2006 eclipse corona (Pasachoff et al. 2007).

A detailed inspection and comparison of our data with those obtained by H. Druckmüllerová from the observing site near Novosibirsk (Russia) with a time lag of 19 minutes showed that the Kreutz family (group I) comet SOHO – C/2008 O1 (Fig. 4) is recorded on wide-angle images taken with a 200 mm telelens. The comet was found the day before eclipse on SOHO C3 images. Our image is the first ground-based eclipse observation of a SOHO comet.

Moreover, at about 10 solar radii northwest of the Sun's center, there is a helmet streamer of a rather unusual form, most likely a remnant of a CME observed by SOHO at this place on the day of the eclipse.

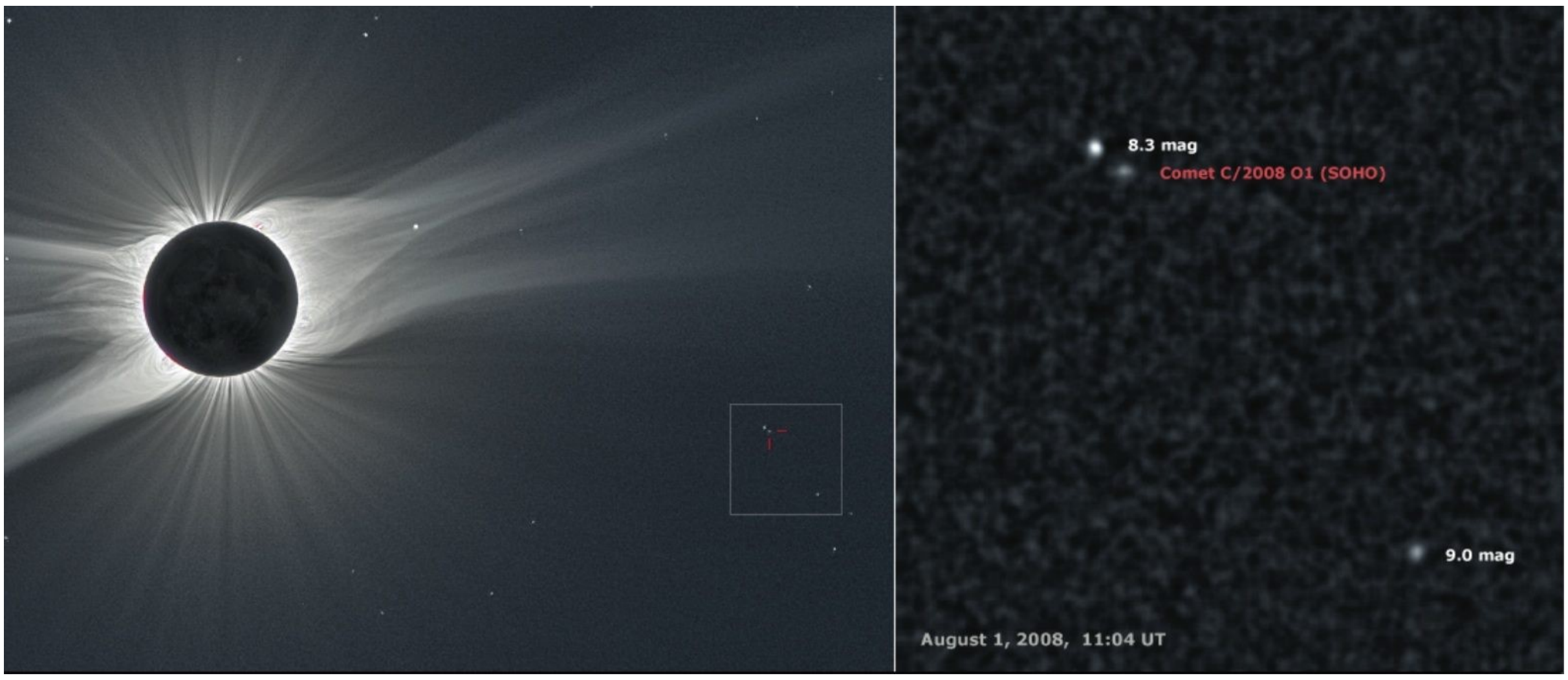

Fig. 4.— An image showing a general view of the white-light corona (left) and the position of the Kreutz family comet C/2008 O1 (right); the two visible stars have magnitudes 8.3 and 9.0, respectively.

## 4. THE GREEN-LINE CORONA

While the white-light corona is essentially the light of the photosphere scattered by free electrons (K-corona), and/or dust particles farther out from the sun (F-corona), the radiation of coronal ions, the emission corona (E-corona), is of emission lines only (e.g., Billings 1966; Golub & Pasachoff 2009). This inherent coronal radiation comes from highly-ionized elements of iron, calcium, nickel etc. A great majority of the corresponding lines, the permitted lines, lie in the extreme UV or X-ray part of the spectrum; in the visible region there are only about 28 lines, all forbidden lines, the brightest of them being Fe XIV at 530.3 nm, whose half-width is about 0.01 nm. For the forbidden lines, the necessary use of narrow-passband filters leads to prolonged exposure times and requires the separation of the line from the continuum.

Some previous observations of the 530.3 nm used a single filter centered on-band (e.g., Badalyan & Sýkora 1997). Since the radial gradient in the intensity of the continuum corona is steep (Koutchmy 1997), the continuum corresponding images must be carefully aligned and subtracted. See also Koutchmy et al. (2005) for discussion of coronagraph and earlier eclipse imaging. At the 2008 eclipse, we simultaneously imaged with two lenses, each with a narrow-passband filter of FWHM 0.15 nm each; one filter was centered at the Fe XIV line itself and the other at the continuum around 529.1 nm (Fig. 5)

From Fig. 5b, given the current period of low emissivity of the [Fe XIV] green-line corona, the main contributor to its intensity is the K-corona. We note that the distributions are confirmed by off-eclipse spectral observations made at the Lomnický štít coronal station; here, the measured intensity at 132° at a height of ~50 arcsec was about 10 ACU, which is a very low value, corresponding to the current solar minimum. (See Rybanský, Rušin, Minarovjech, Klcok, and Cliver, 2005, for a discussion of the coronal index of solar activity and the definition of Absolute Coronal Units, ACU, and Badalyan and Obridko, 2006, for further discussion of ACU as a millionth of the brightness at disk center in a 0.1-nm band of

the adjacent continuum.)

In the infrared coronal forbidden emission line [Fe XI] 798.2 nm, Habbal et al. (2007) imaged to at least 3 solar radii at the 2006 eclipse.

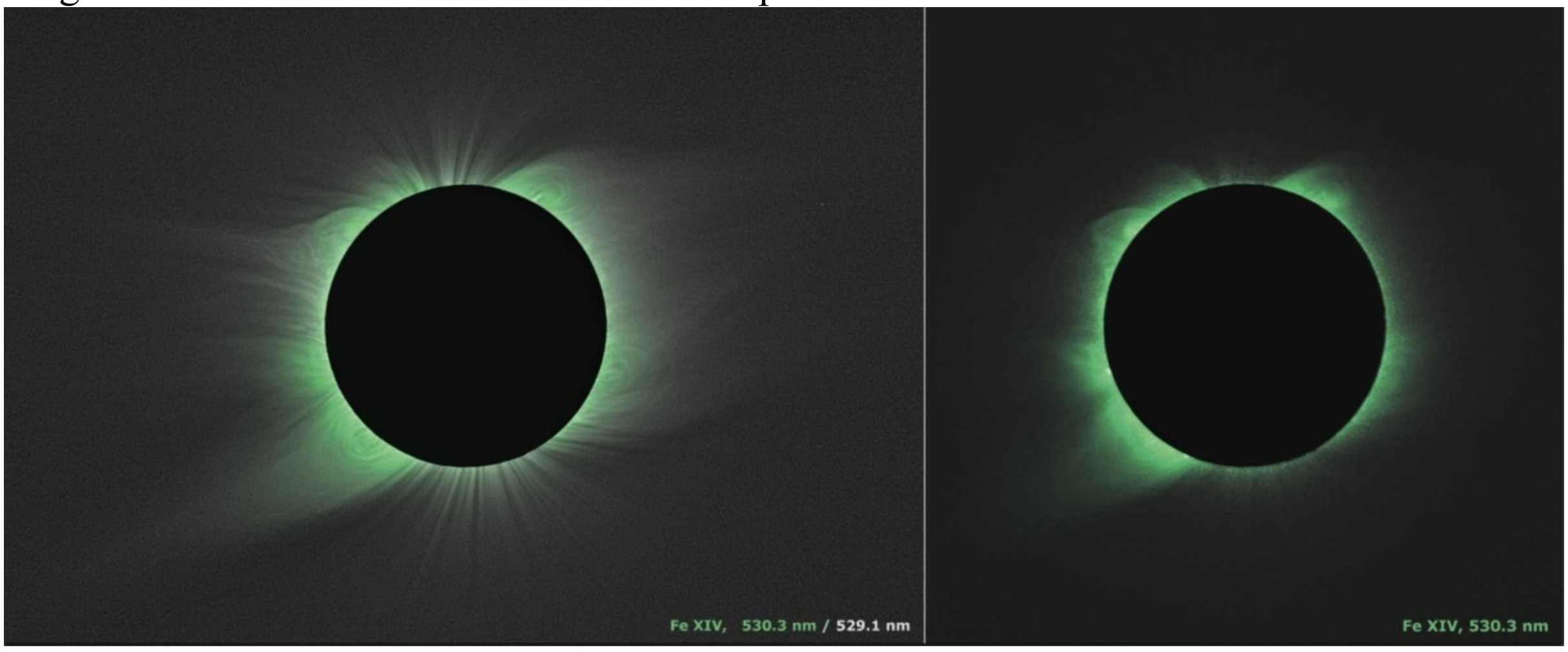

(a)

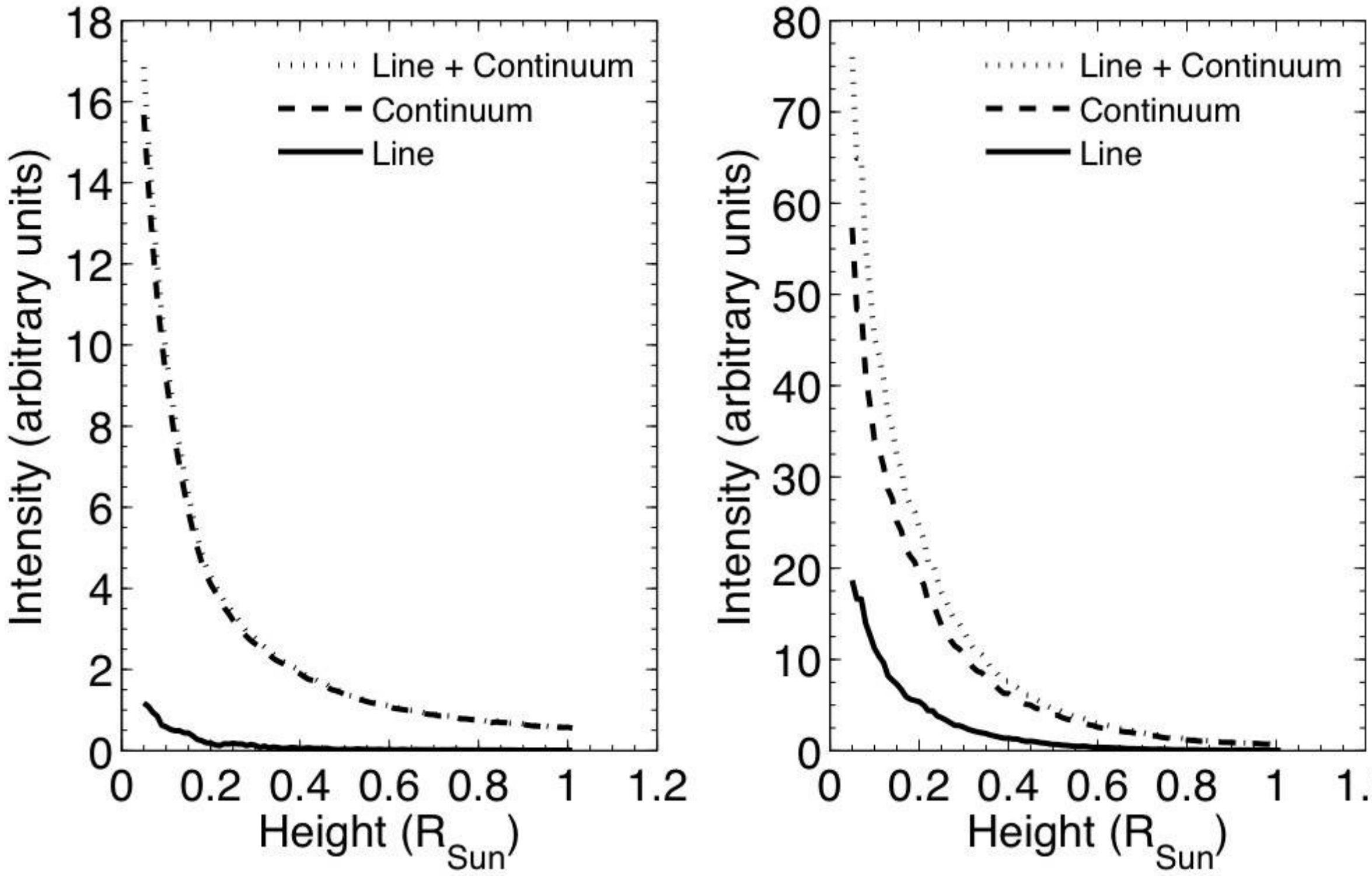

Fig. 5.— (a) Subtracting a continuum image on band from the composite of the continuum- and green-light coronas (left), gives an image in the "pure" green-line corona (right), which shows the distribution of [Fe XIV]. (b) The radial distributions of the intensities of the coronal green line ([Fe XIV], 530.3 nm) and the continuum, the continuum in the closer neighborhood of the line and their difference. The radial scans were made at position angles of 0° (left) and 132° (right).

## 5.— COMPARISON BETWEEN SITES

Our sites in Mongolia and Siberia were separated by 19 minutes in travel time for the umbra (Espenak and Anderson, 2007). Even with such a small time difference, the high contrast and high resolution of our image-processing allows many motions in the corona to be detected (Figs. 6, 7, and 8).

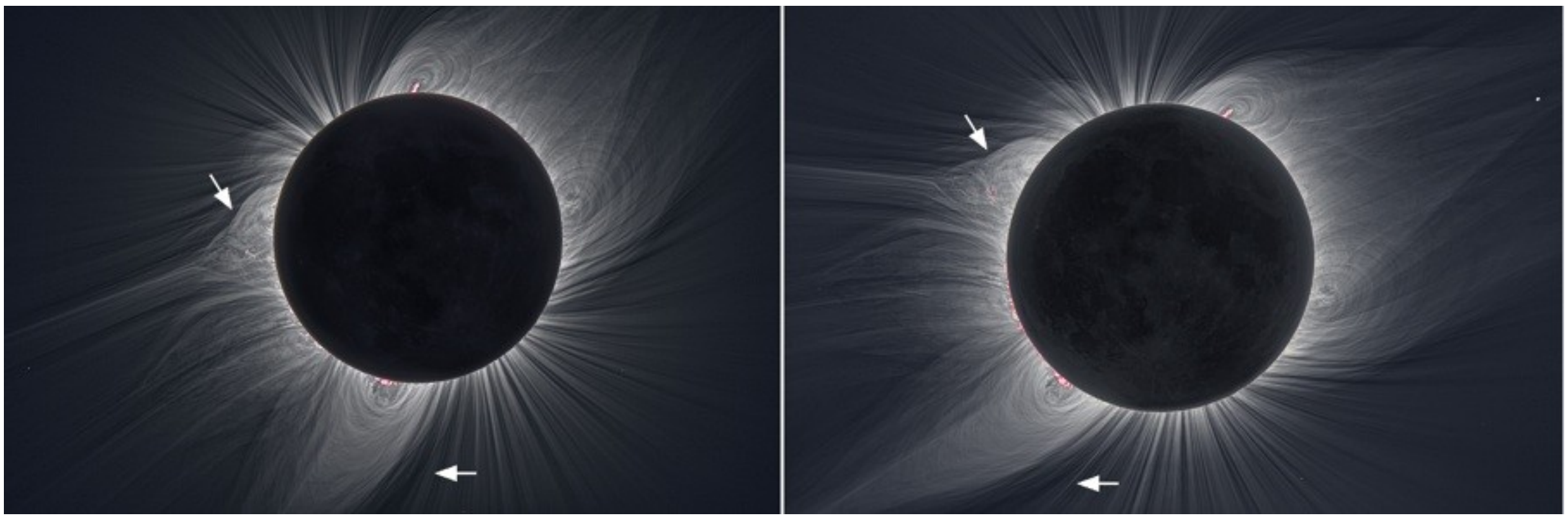

Fig. 6.— Comparison of images separated by 19 min in time, from Novosibirsk, Russia, to Altaj, Mongolia, shows a variety of coronal and prominence motions. Several regions show remarkable small-scale changes. Note also the lunar features illuminated by earthshine. An open research question, suggested by Anthony Mallama, is whether the lunar albedo peaks at 0° phase angle, and our observations may prove useful for resolving this question.

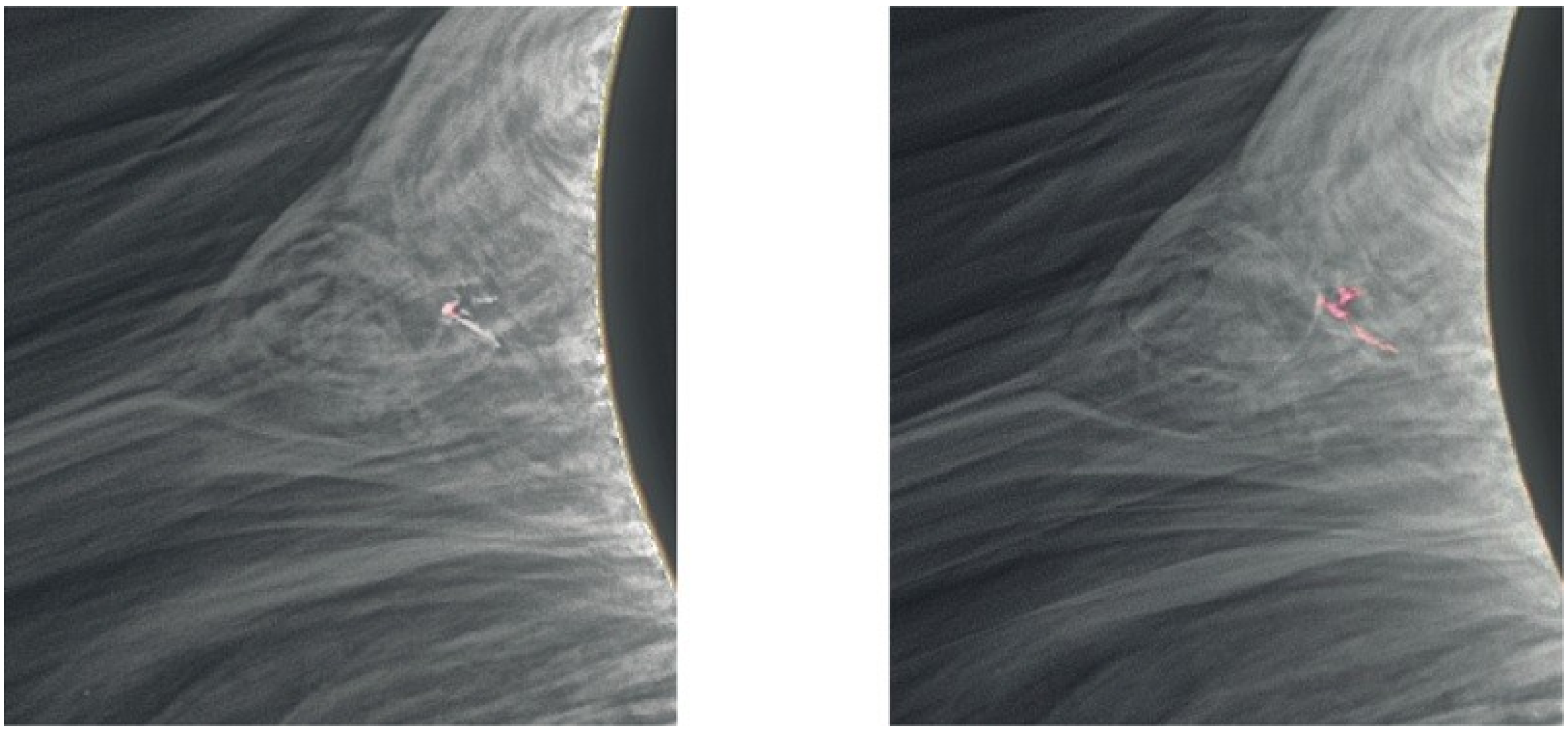

Fig. 7.—Comparison showing motions on the east limb between images separated by 19 min of time between Novosibirsk and Altaj. Changes in the fine structure of the region are remarkable.

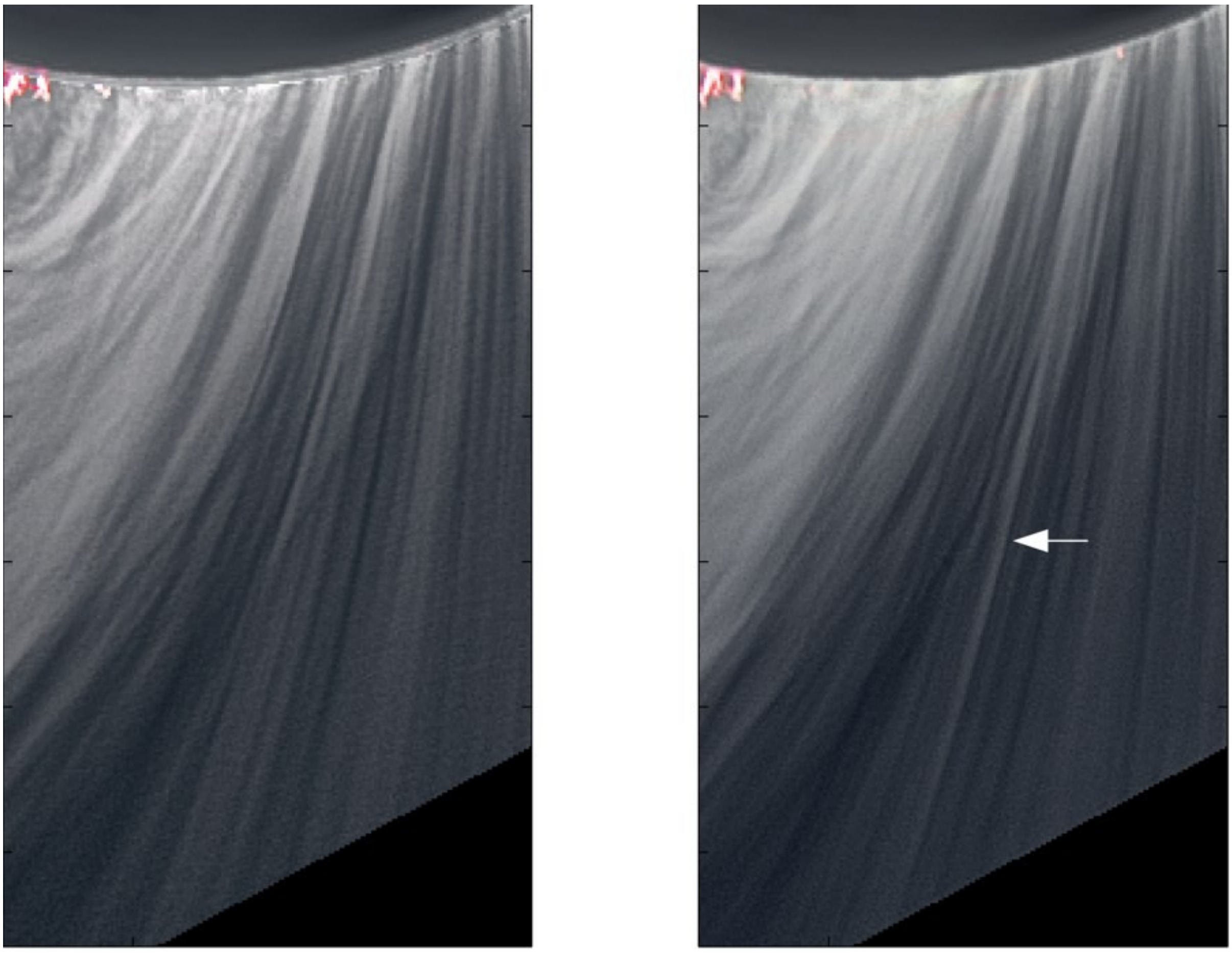

Fig. 8.— An illustration of the dynamics of the corona above the south pole. A new polar plume is indicated by an arrow. An estimated speed of the ejection is ~600 $km^{-1}$.

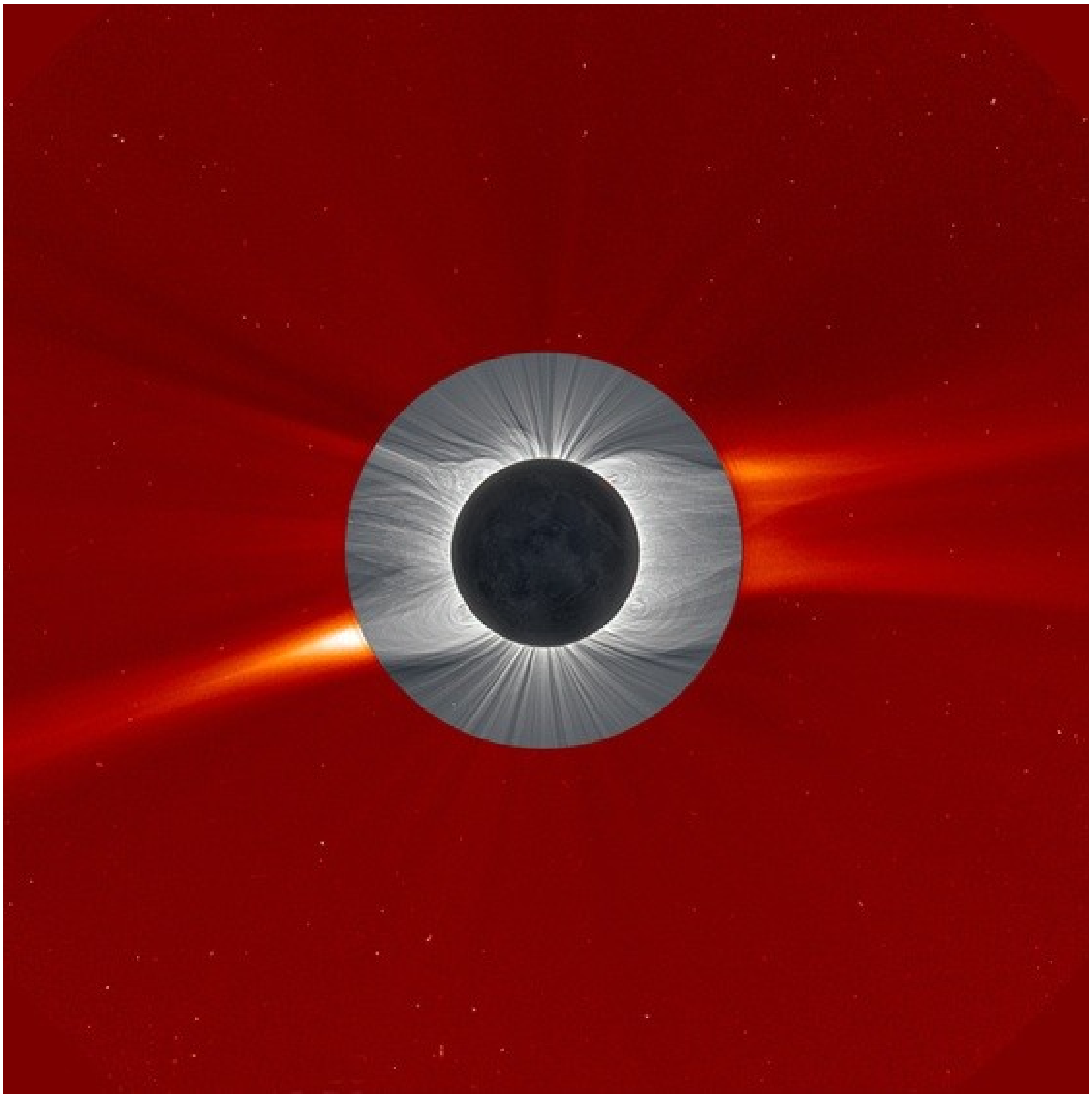

111

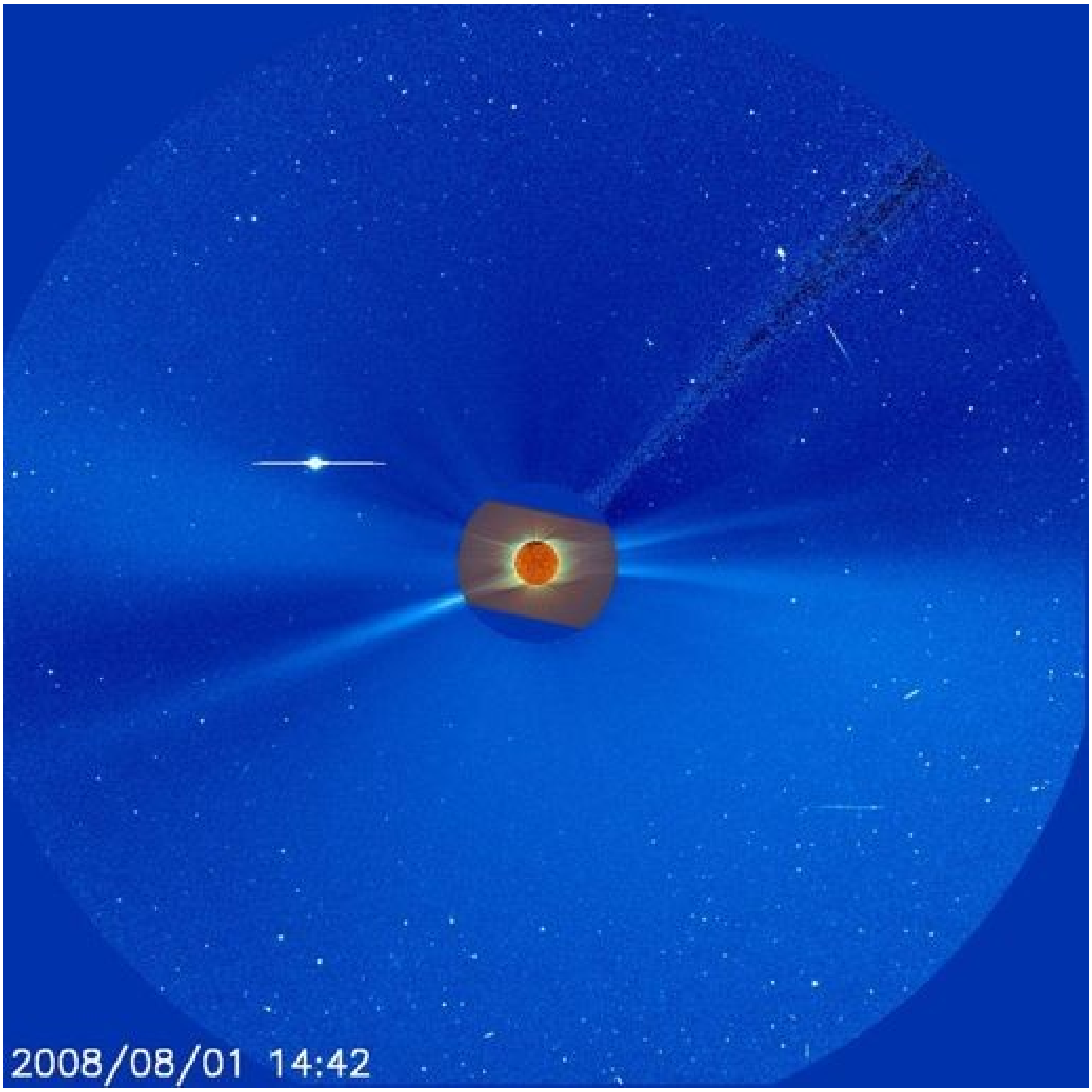

Fig. 9.— (a) A composite showing our eclipse image from Mongolia inserted in the missing lower-coronal region that is inaccessible to the SOHO C2 coronagraph. Note how sparse, and limited to equatorial latitudes, the streamers were at this low phase of the solar-activity cycle. Mercury, with saturated pixels to each side, is visible at magnitude –1.7 and solar elongation of 3.4°. The silhouette of the rod that holds the occulting disk in place shows diagonally to upper right from the Sun to the edge of the grame. (b) A composite with our eclipse image inserted in the occulted region of the SOHO C3 coronagraph. (Courtesy of Steele Hill, NASA Goddard Space Flight Center)

**6. COMPARISON WITH SOHO IMAGING**

The ESA/NASA Solar and Heliospheric Observatory (SOHO) was carrying out an [Fe XII] 19.5 nm campaign with constant reimaging at that wavelength during the hours surrounding the eclipse and during the eclipse itself. It obtained a full set of Extreme-ultraviolet Imaging Telescope (EIT) images several hours later. Its Large-Angle Spectrometric Coronagraph (LASCO) C2 and C3 images have void doughnuts of coverage, which we fill in on eclipse days (Fig. 9). The data-processing methods reported in this paper recover coronal structure sufficiently far out that they substantially overlap the LASCO coronagraphs. On the solar disk, the quiet-sun configuration left few obvious active features to which one could trace the bases of the streamers that show on the eclipse images; of course, half the bases are on the invisible, far side of the Sun.

## 7. COMPARISON WITH STEREO IMAGING

The perspectives of the corona from the twin STEREO spacecraft, from approximately 30° in front of and behind the Earth in its solar orbit on eclipse day, give very different views of the streamers on the Sun at eclipse time (Fig. 10).

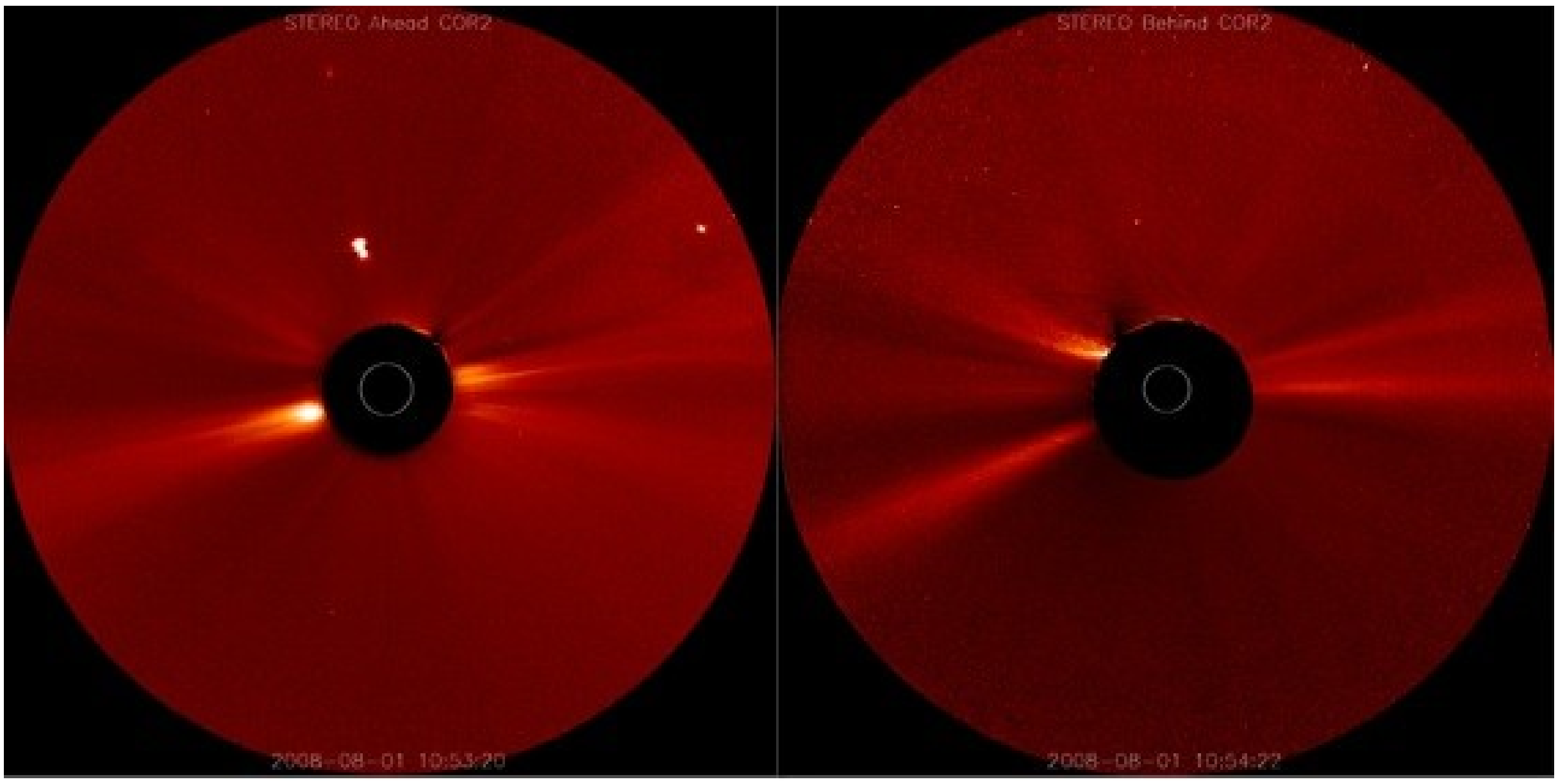

Fig. 10.— The images from NASA's STEREO pair of spacecraft that correspond to eclipse time. (left) A (Ahead) was 34.9° ahead the Sun-Earth line and (right) B (Behind) was 29.6° behind with 64.5° between them. The size of the solar photosphere is drawn in as a white circle on the regions hidden by the occulting disks. The STEREO A image shows Venus at magnitude –3.9. Semidiameters: B=886.676, Earth=945.498, and A=1002.248. (Courtesy of Steele Hill, NASA Goddard Space Flight Center; STEREO team, NRL; and NASA)

## 8. THE CORONA AND THE SOLAR-ACTIVITY CYCLE

The 2008 eclipse occurred at the lowest point in the solar-activity cycle in many decades (Fig. 11). No sunspots were on the sun at all—sunspot number = 0—for more than half the days in 2008. We calculated the Ludendorff flattening coefficient to be 0.29, which corresponds well to the graph of previous measurements (Koutchmy & Nitshelm 1984), and which we hereby also update for 2001's eclipse (0.165) and 2006's eclipse (0.170) (Rušin 2000; Özkan et al. 2007) from measurements on Druckmüller's images.

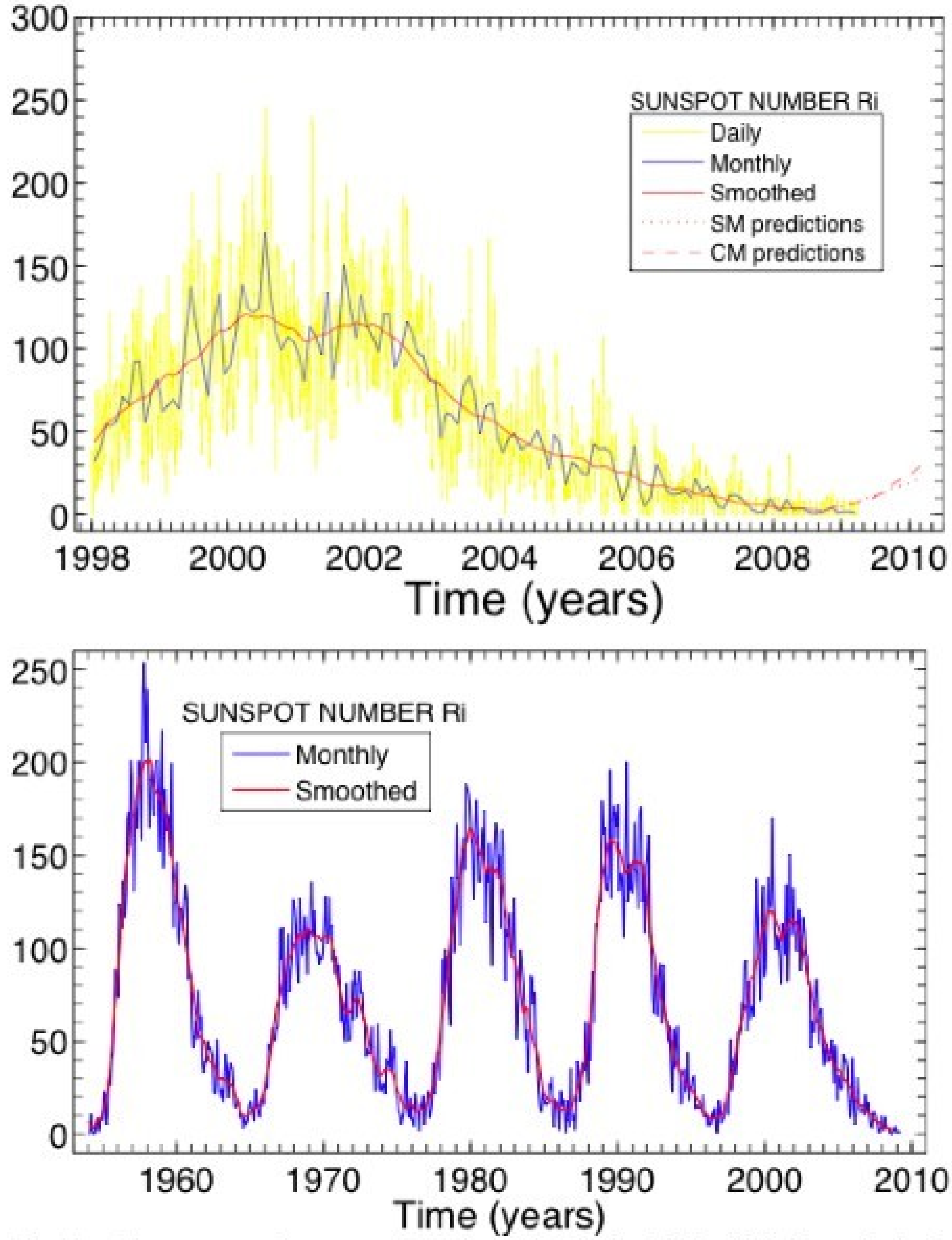

Fig. 11.— The sunspot cycle was at a many-decade low during 2008, which shows clearly in the coronal configuration at the eclipse. (a) The most recent cycle. (b) The most recent five cycles, none of which averaged as low as the current minimum. (courtesy of Laurence Wauters, Solar Influences Data Analysis Center, Royal Observatory of Belgium)

The absence of activity on the solar disk showed clearly in the ground-based magnetic field maps (Fig. 12) and in the MDI (Michelson Doppler Imager on SOHO) accumulated view of the solar surface (Fig. 13).

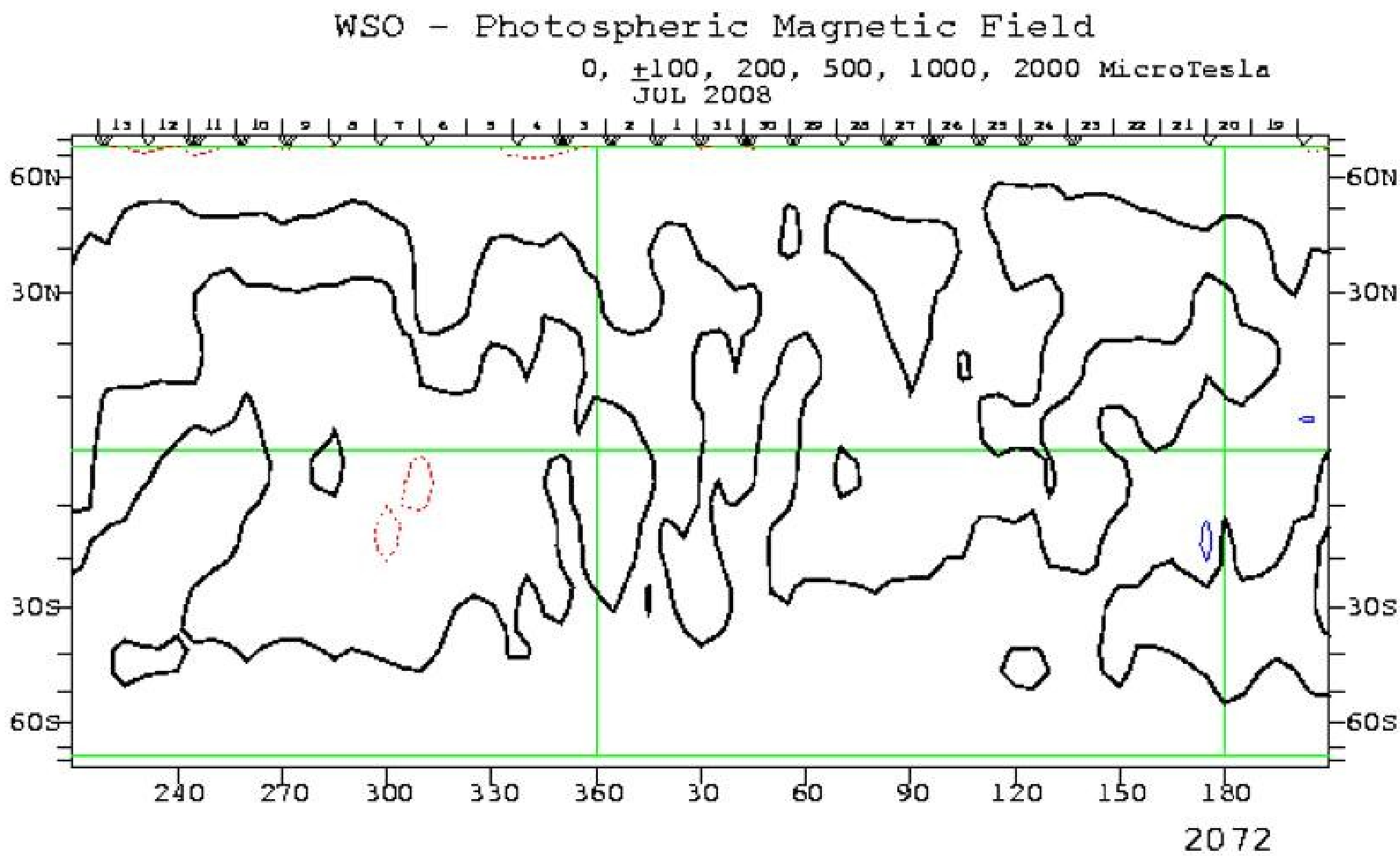

Fig. 12.— The photospheric magnetic field, mapped at the Wilcox Solar Observatory, from daily vertical strips at the central meridian. Carrington Rotation 2072 ranged from 2008 July 6 to 2008 August 3; the diagram shows the magnetic field centered on eclipse day. (courtesy of Todd Hoeksema and Philip Scherrer, Stanford University)

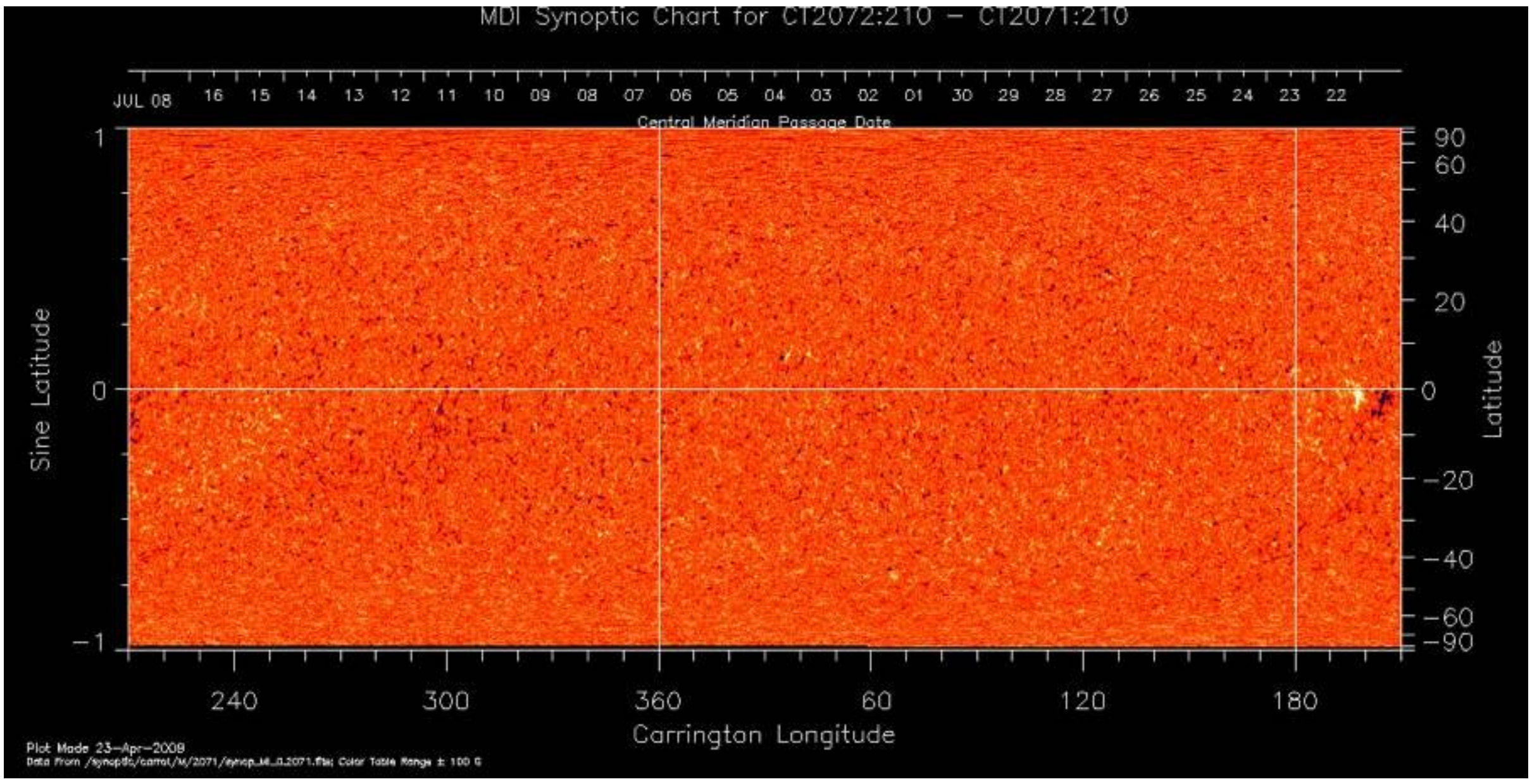

Fig. 13.— The magnetic-field view of the solar surface from the Michelson Doppler Imager (MDI) on SOHO, centered on eclipse day, unrolled. (courtesy of Todd Hoeksema & Philip Scherrer, Stanford University)

The extremely low stage of the solar-activity cycle also shows the STEREO-B ultraviolet images (Fig. 14) and in the TRACE ultraviolet images (Fig. 15). We show the base of the plumes in the north coronal hole from SOT (Fig. 16). The low phase of the solar-activity cycle also shows clearly in the Hinode XRT x-ray image (Fig. 17).

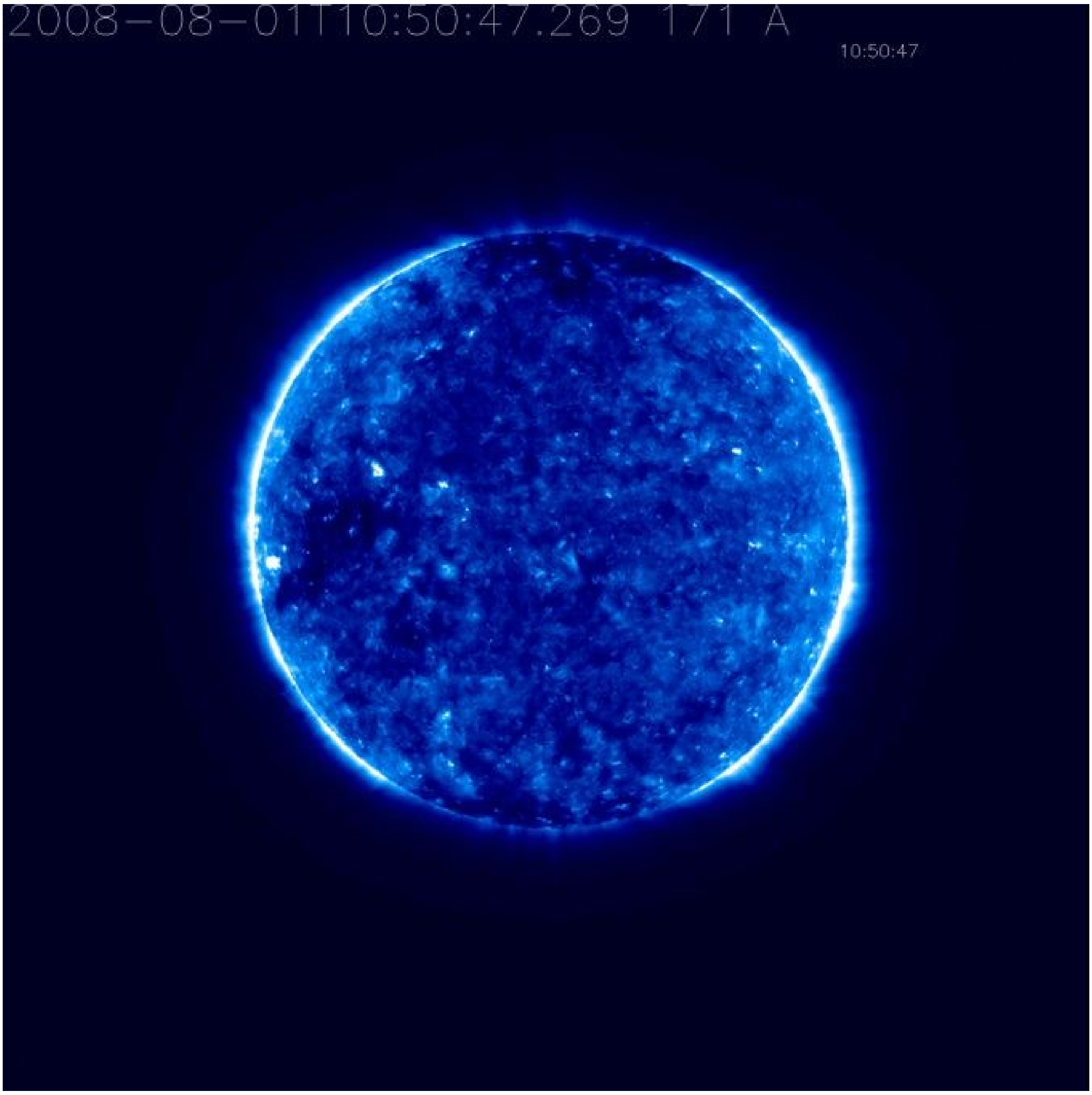

(a)

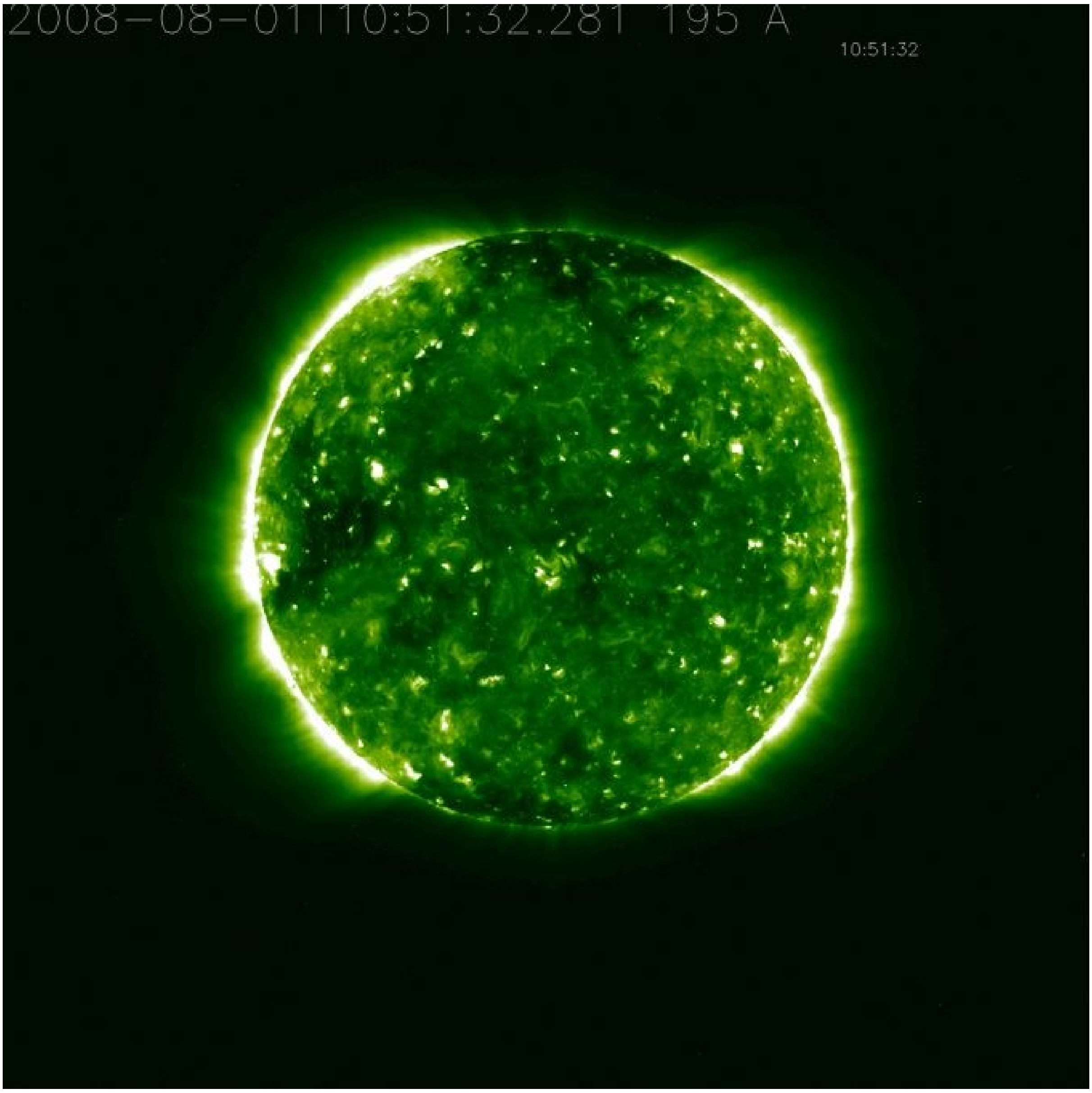

(b)

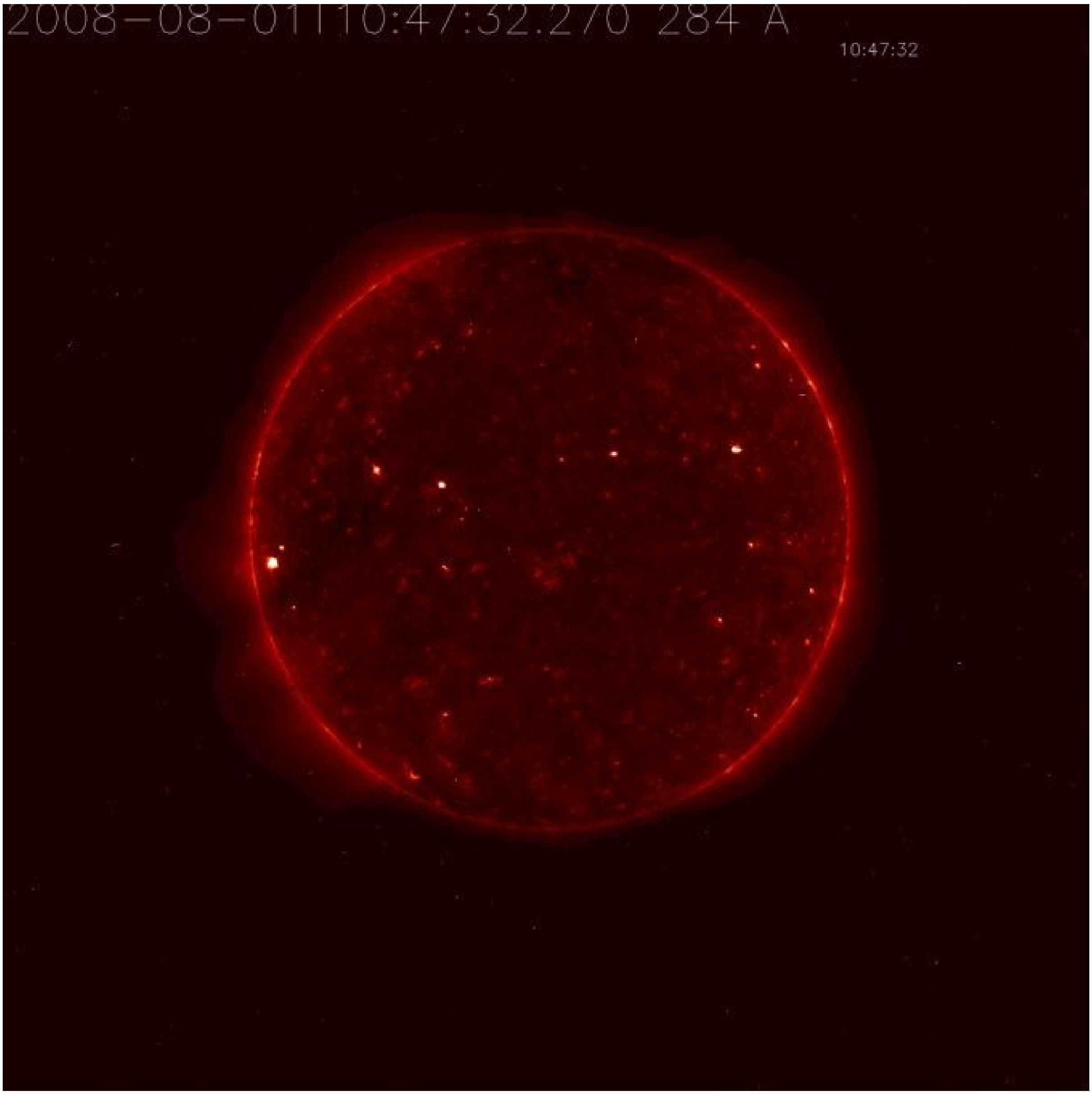

(c)

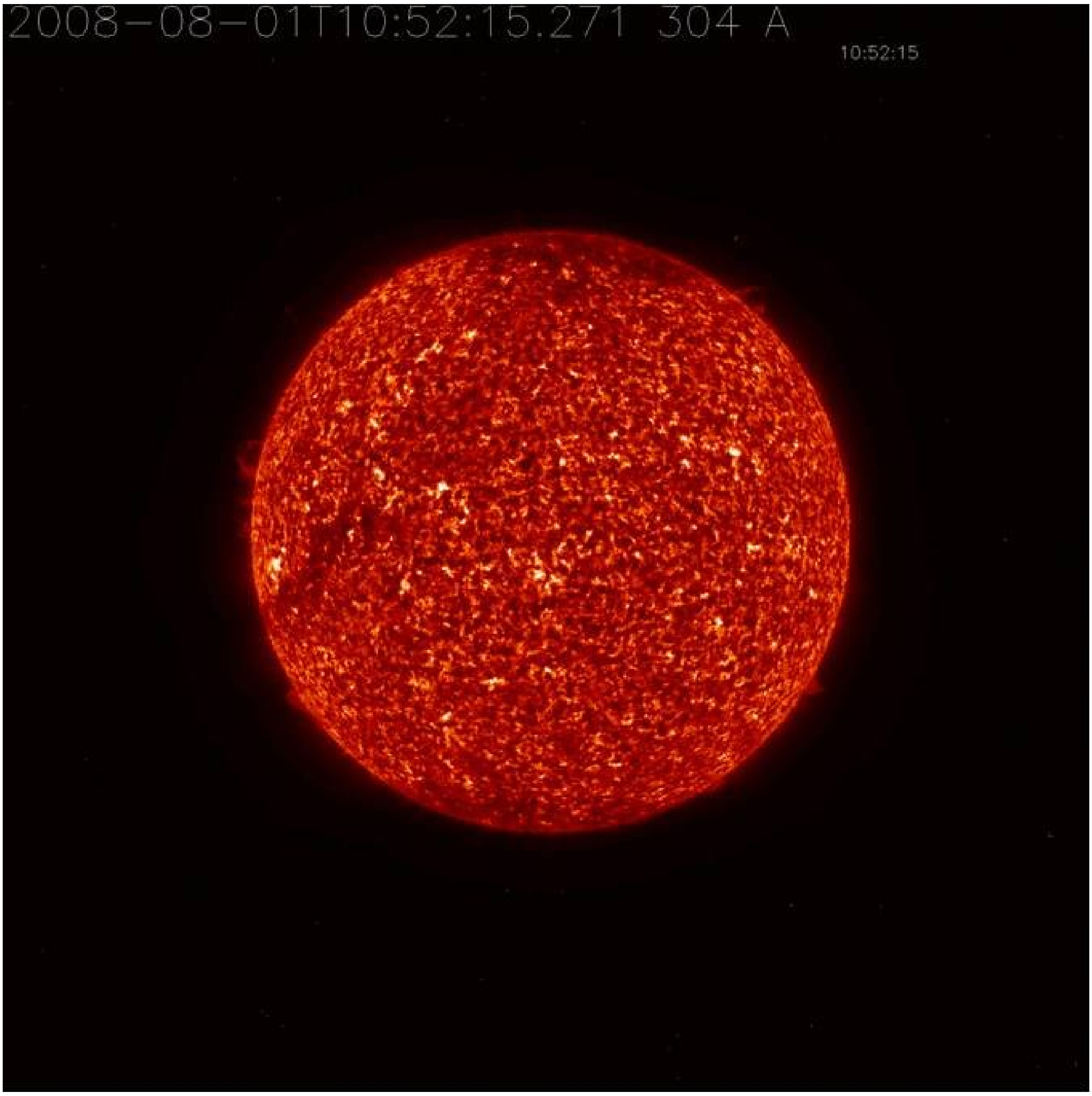

(d)

Fig. 14.— The STEREO-B EUV imager shows hardly any activity anywhere on the disk and only small coronal holes at the poles. (a) 17.1 nm ([Fe IX]); (b) 19.5 nm ([Fe XII); (c) 284 nm ([Fe XV]); and (d) 304 nm (He II); see Golub & Pasachoff (2009).

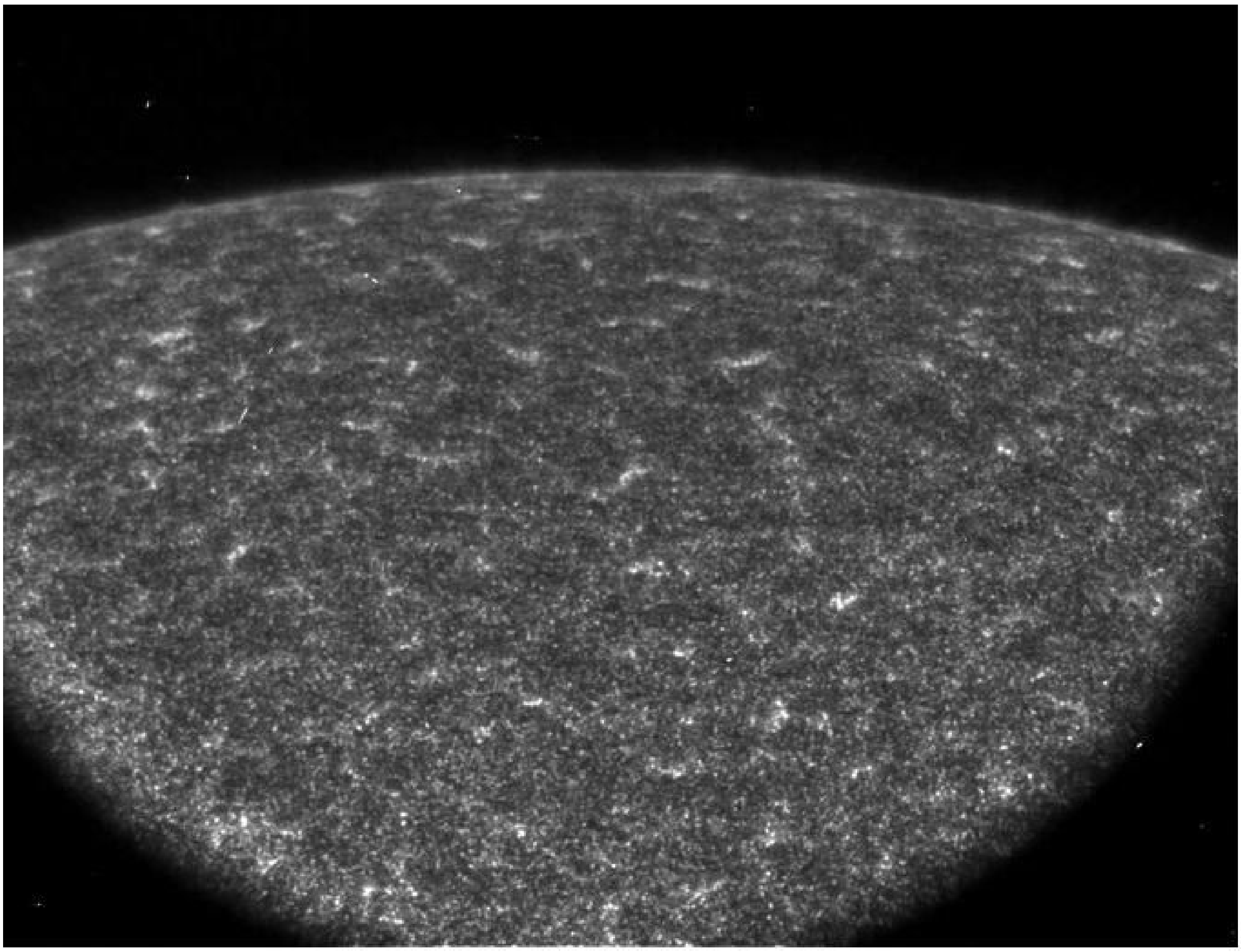

Fig. 15.— The continuum image at the north polar limb at 160 nm from NASA's Transition Region and Coronal Explorer (TRACE) shows no activity.

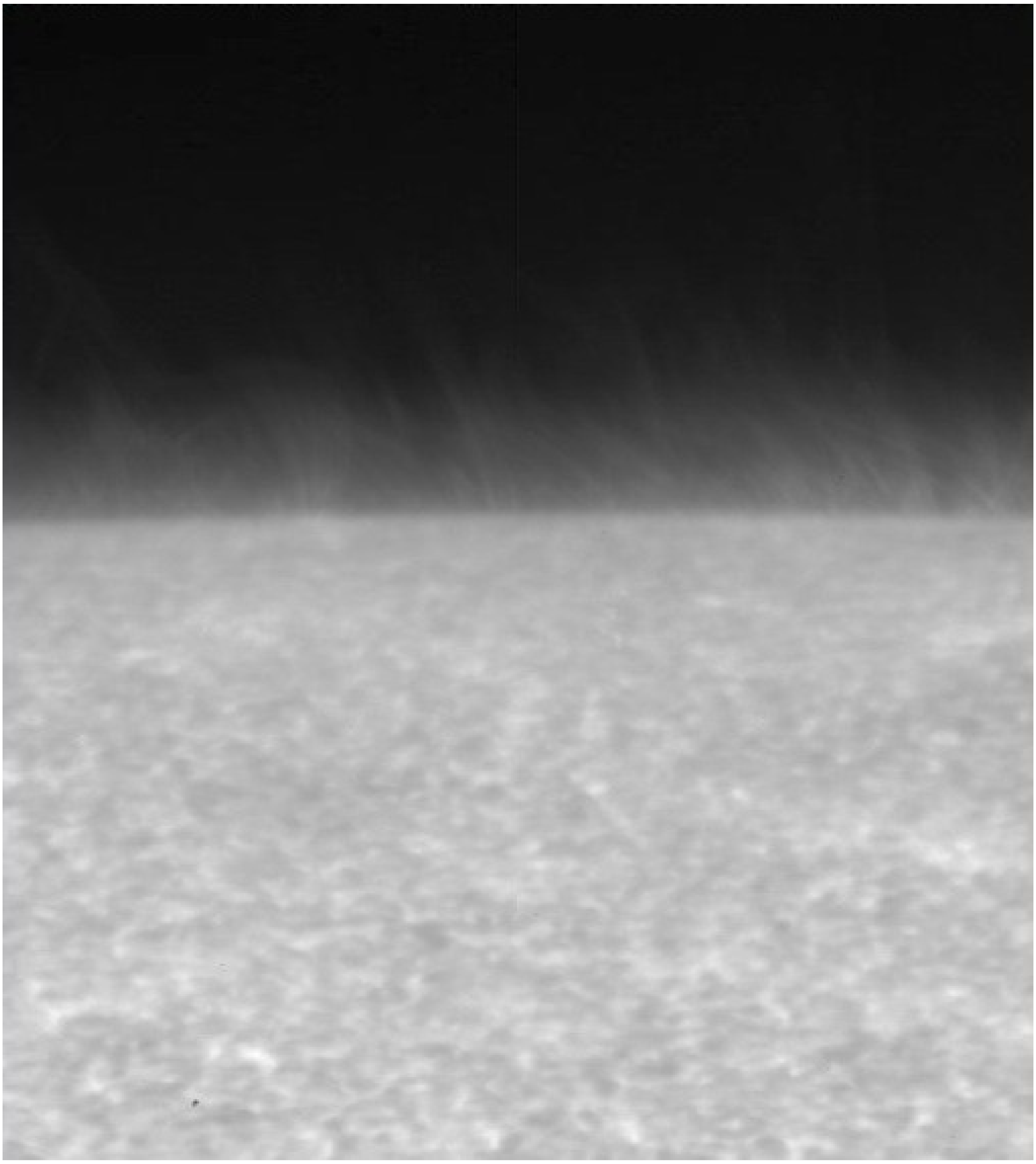

Fig. 16.— A north polar image in the Ca II H line, one taken from a series at 30 s cadence and 55×110 FOV by SOT on Hinode. Since the filter is 0.25 nm wide, the image shows mostly upper photospheric light on the disk but off the limb it shows the integrated chromospheric emission. Longer exposures in the future may help in providing the possibility of aligning these chromospheric features with coronal features. (Courtesy of Ted Tarbell, Lockheed Martin Solar and Astrophysics Laboratory)

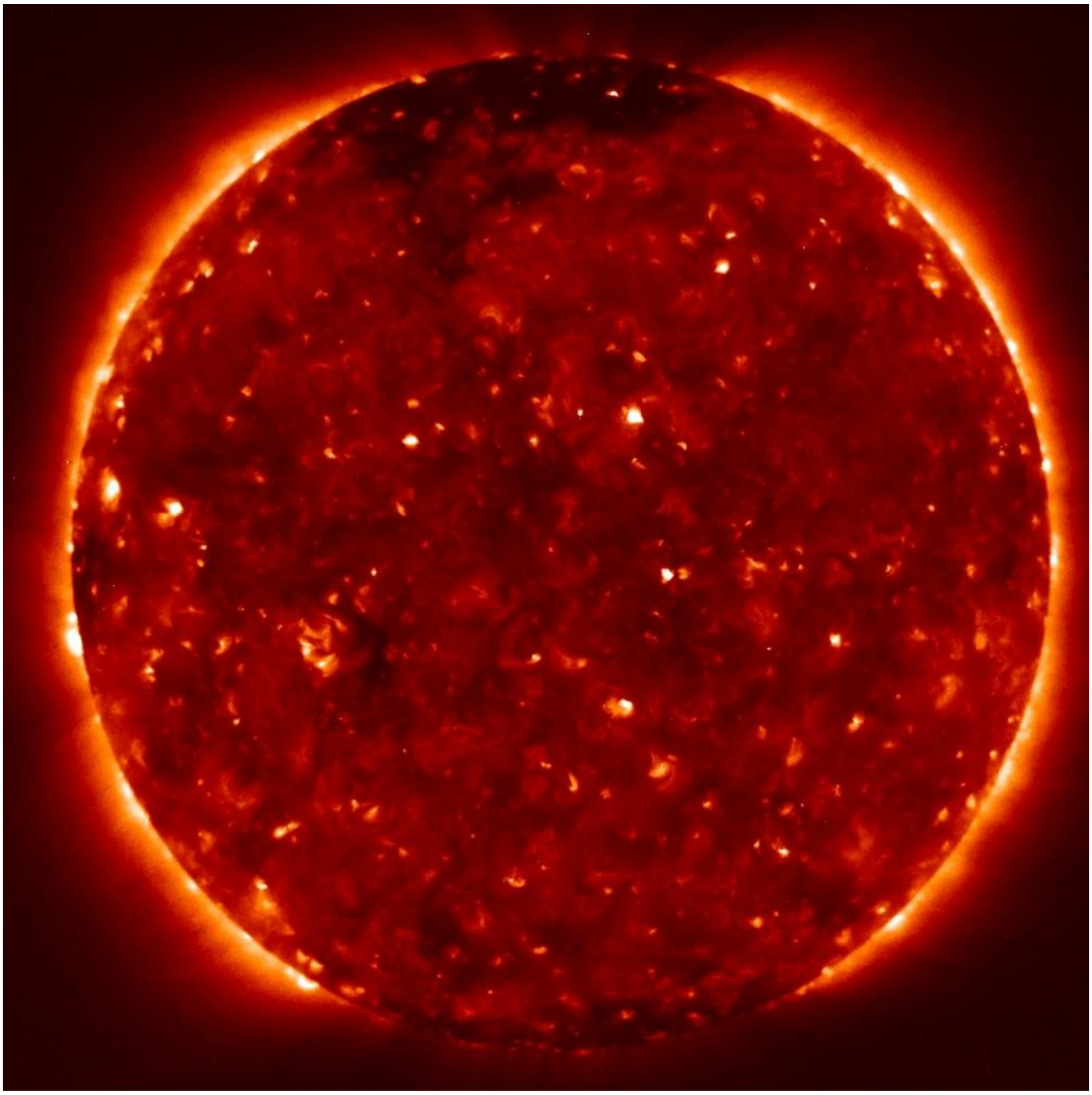

(a)

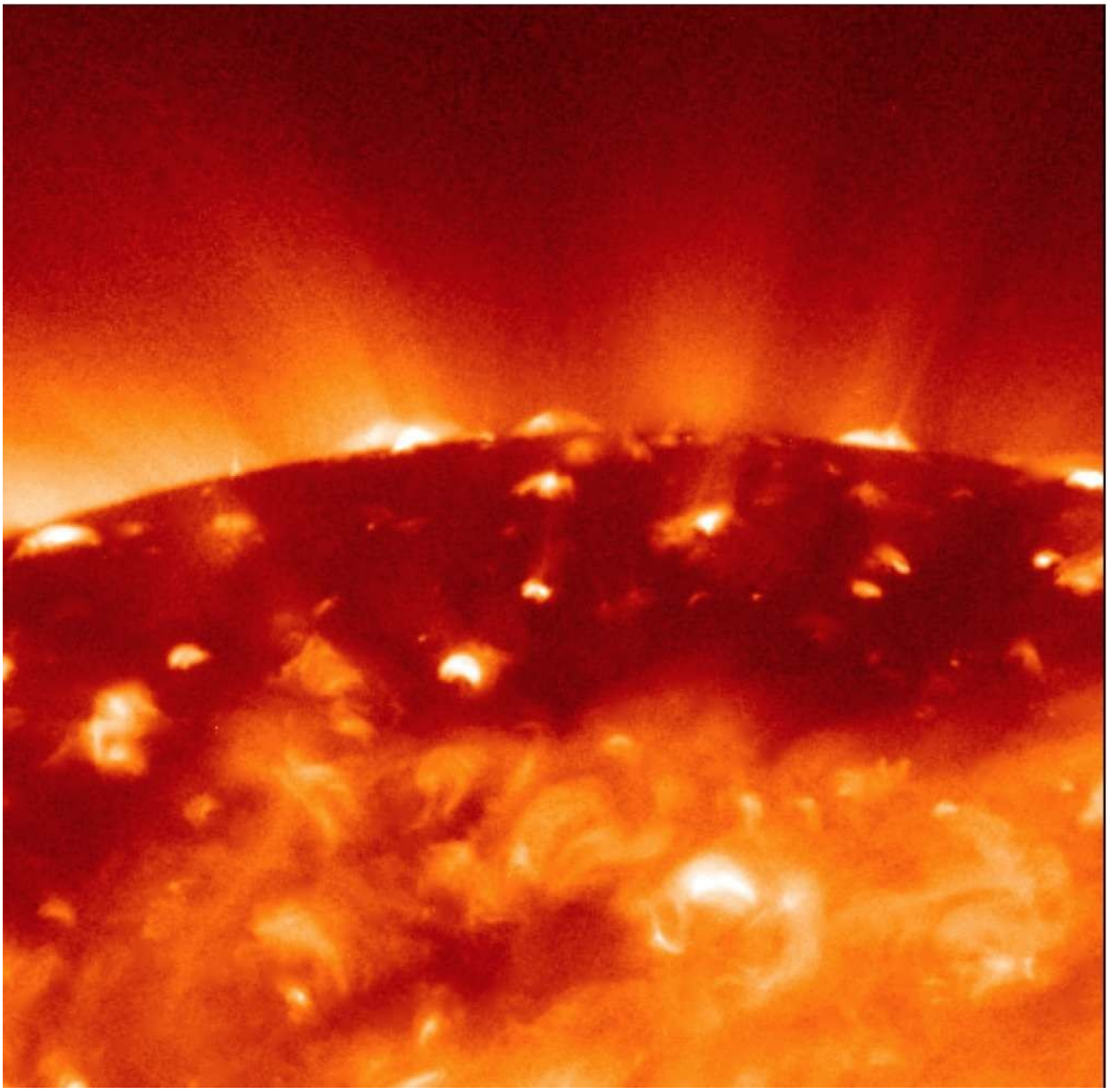

(b)
Fig. 17.— X-ray images from the X-Ray Telescope (XRT) on Hinode. (a) The full-disk (synoptic) image at 10:05 UT; the synoptic image closest to our totality was at 10:16 UT. (b) A north-polar image at 10:58 UT. (Courtesy of Leon Golub and Alexander Engell, Harvard-Smithsonian Center for Astrophysics)

## 9. CONCLUSIONS

Our preliminary analysis shows that the 2008 August 1 total eclipse white-light corona possesses a highly-complex and intricate structure, especially above the eastern limb of the Sun. Its large-scale shape agrees well with the one predicted by Mikic et al. (2008) based on a magnetohydrodynamic model of the solar corona with improved energy transport. The green-light corona is weak, which is typical for periods around minima of solar cycles.

An occurrence and particular distribution of helmet streamers indicate that a new solar cycle has already begun, despite only a very sporadic occurrence of sunspots. Identification of the comet C/2008 O1 and a peculiar CME remnant as observed by SOHO shows that a variety of novel small-scale features are real objects in the corona rather than artifacts of the data processing method employed. These fine structures not only pose a serious challenge for serious theoretical modelling of the processes shaping the corona, but should also be taken into account in preparation for future ground-based eclipse observations of the corona, in particular those of the 22 July 2009 and 11 July 2011 total eclipses (Pasachoff 2009b).

We thank Peter Aniol and Martin Dietzel for their assistance with eclipse photography. This work (Rušin, Saniga) was partially supported by the Slovak Academy of Sciences Grant Agency VEGA, Grant 7012, and by the Science and Technology Assistance Agency APVT under contract APVT 51-012-704. Druckmüller's research was partially supported from the the Czech Science Foundation, grant GACR 205/09/1469/2. Aniol acknowledges the support received from ASTELCO. The identification of the SOHO comet was possible thanks to private communication from M. Kusiak, M. Mašek and D. Hanžl. Our special thanks go to Prof. Batmunch of the Center for Astrophysics and Geophysics, Ulaan Baattar, for his kind guidance, help and assistance at various stages of the expedition. Finally, we are grateful to the other members of our team – M. Ditzer, Ľ. Klocok, K. Martišek, J. Sladeček, and P. Zimmermann – who substantially contributed to the success of the expedition to Mongolia. For help with the spacecraft-data figures, we thank Leon Golub and Alexander Engell of the Harvard-Smithsonian Center for Astrophysics and Steele Hill of NASA's Goddard Space Flight Center. For help with the magnetic field maps from the ground and from space, we thank Todd Hoeksema and Philip Scherrer, Stanford University. Pasachoff's eclipse work is supported in part by the Committee for Research and Exploration of the National Geographic Society. He thanks the Planetary Sciences Department of the California Institute of Technology for sabbatical hospitality. We are indebted to Prof. Mark Stuckey (Elizabethtown College, PA) for careful proofreading of an earlier version of the paper. The Transition Region and Coronal Explorer, TRACE, is a mission of the Stanford-Lockheed Institute for Space Research (a joint program of the Lockheed Martin Advanced Technology Center's Solar and Astrophysics Laboratory and Stanford's Solar Observatories Group), and part of the NASA Small Explorer program. XRT's telescope, a joint project of the Smithsonian Astrophysical Observatory and Japanese partners, is aloft on Hinode, a mission developed and launched by ISAS/Japan Aerospace Exploration Agency (JAXA), with the National Astronomical Observatory of Japan as domestic partner and NASA and STFC (UK) as international partners. It is operated by these agencies in co-operation with ESA and the NSC (Norway). The SOT image was provided by Ted Tarbell of LMSAL, supported by NASA contract NNM07AA01C.